\documentclass[twocolumn,superscriptaddress]{revtex4-2}

\usepackage{graphicx}
\usepackage{dcolumn}
\usepackage{bm}
\usepackage{mathtools}
\usepackage{amssymb}
\let\vec\mathbf
\usepackage{physics}
\usepackage{xcolor}
\usepackage[utf8]{inputenc}
\usepackage[T1]{fontenc}
\usepackage{mathptmx}
\usepackage{etoolbox}
\usepackage{placeins}
\usepackage{algorithm}
\usepackage{algpseudocode}
\usepackage{multirow}
\usepackage{booktabs}
\makeatletter
\def\@email#1#2{%
 \endgroup
 \patchcmd{\titleblock@produce}
  {\frontmatter@RRAPformat}
  {\frontmatter@RRAPformat{\produce@RRAP{*#1\href{mailto:#2}{#2}}}\frontmatter@RRAPformat}
  {}{}
}%
\makeatother

\begin{document}



\title{Excited States from Restricted Open Shell Plane-Wave DFT}

\author{Michael J. Sahre}
\thanks{michael.sahre@univie.ac.at}
\affiliation{University of Vienna, Faculty of Physics, Kolingasse 14, A-1090 Vienna, Austria}

\author{Marco Romanelli}
\affiliation{University of Vienna, Faculty of Chemistry,  Institute of Theoretical Chemistry, \\ W\"ahringer Str. 17, 1090 Vienna, Austria}

\author{Martijn Marsman}
\affiliation{VASP Software GmbH, Berggasse 21, A-1090 Vienna, Austria}

\author{Leticia Gonz\'{a}lez}
\affiliation{University of Vienna, Faculty of Chemistry,  Institute of Theoretical Chemistry, \\ W\"ahringer Str. 17, 1090 Vienna, Austria}

\author{Georg Kresse}
\affiliation{University of Vienna, Faculty of Physics, Kolingasse 14, A-1090 Vienna, Austria}
\affiliation{VASP Software GmbH, Berggasse 21, A-1090 Vienna, Austria}

\begin{abstract}
Variational excited-state density functional theory (DFT) enables the calculation of excited states at a cost comparable to ground-state calculations, but single-configuration approaches often suffer from spin contamination. We implement restricted open-shell Kohn-Sham (ROKS) DFT, which recovers spin-pure singlet excitation energies via the variational minimization of a weighted combination of mixed-spin and triplet configurations, within the plane-wave projector augmented-wave framework of VASP. The energy functional is optimized using a preconditioned conjugate-gradient or a direct inversion in the iterative subspace algorithm, and analytical atomic forces are derived.
The implementation is validated for eight organic molecules by comparison to the Q-Chem quantum chemistry code, yielding mean deviations of approximately $30\,\mathrm{meV}$. As a solid-state application, we investigate the three lowest lying excitations of MgO with a neutral oxygen vacancy. For a dielectric-dependent hybrid functional, vertical excitation energies from ROKS and time-dependent density functional theory (TDDFT) differ on average by about $0.21\,\mathrm{eV}$. The Franck-Condon shifts deviate on average by 0.14~eV between the two methods and mass-weighted displacements between the excited states and the ground state by 0.12~amu$^{1/2}$\AA. Additional calculations at the PBE level reveal that these properties depend less strongly on the DFT functional for ROKS than for TDDFT.
These results demonstrate that ROKS provides excitation energies and excited-state forces with an accuracy similar to TDDFT while retaining the favorable scaling of ground-state DFT, making it a promising approach for affordable excited-state simulations in extended systems.
\end{abstract}

\maketitle 

\section{Introduction}
Modeling excited state properties can provide valuable information for designing materials for optoelectronic devices such as light-emitting diodes or for applications like solar energy conversion.
However, a challenge in the modeling of relevant materials is the substantial number of atoms necessary for a realistic representation of systems such as defective bulk materials or surfaces. This necessitates an excited state method that is computationally inexpensive, suitable for applications to systems comprising hundreds or thousands of atoms. Furthermore, the method should provide excited state forces (nuclear gradients) if one wishes to investigate relaxation processes in excited states.

A method that incorporates these features is $\Delta$-self-consistent field ($\Delta$-SCF)~\cite{SLATER19721,Ziegler1977,PhysRevB.78.075441} density functional theory (DFT), which is also referred to as excited state~\cite{Levi2020} or orbital-optimized DFT~\cite{Hait2021}.
In this approach, an excited electronic configuration of the ground state is created by promoting an electron to an unoccupied orbital. The orbitals of this excited configuration are then relaxed under the constraint that the system remains in this excited configuration. The corresponding excitation energy can then be calculated from the difference between the excited state and the ground state energy. Additionally, forces can be evaluated via the same formalism as for the ground state since the excited state is stationary with respect to the energy.
A specific excited state can hence be calculated at the cost of a ground state calculation.
Consequently, the method has been used to study the optical properties of point defects in various semiconductors, which requires large supercells to suppress interactions between periodic images.
For instance, the excitation energies, zero phonon lines, hyperfine tensors and photoluminescence spectra have been calculated for different materials, such as the nitrogen vacancy in diamond~\cite{PhysRevB.77.155206,Alkauskas_2014,Ivanov2023}, point defects in silicon carbide polymorphs~\cite{PhysRevB.96.085204,PhysRevMaterials.7.096202,https://doi.org/10.1002/adma.202408424}, and hexagonal and rhombohedral boron nitride~\cite{PhysRevLett.123.127401,Cholsuk2024,https://doi.org/10.1002/adom.202500593}.

Nevertheless, a disadvantage of the $\Delta$-SCF method is that the desired excited configuration may collapse to the ground state or a lower-lying excited state during the optimization procedure. Therefore, a variety of techniques have been developed to circumvent such a collapse. 
These include for example the (initial) maximum overlap method~\cite{doi:10.1021/jp801738f,doi:10.1021/acs.jctc.7b00994}, the state targeted energy projection~\cite{doi:10.1021/acs.jctc.0c00502}, squared gradient minimization~\cite{Hait2020} or a generalized eigenmode following approach~\cite{Schmerwitz2023}.

\begin{figure}
    \centering
    \includegraphics[width=1.0\linewidth]{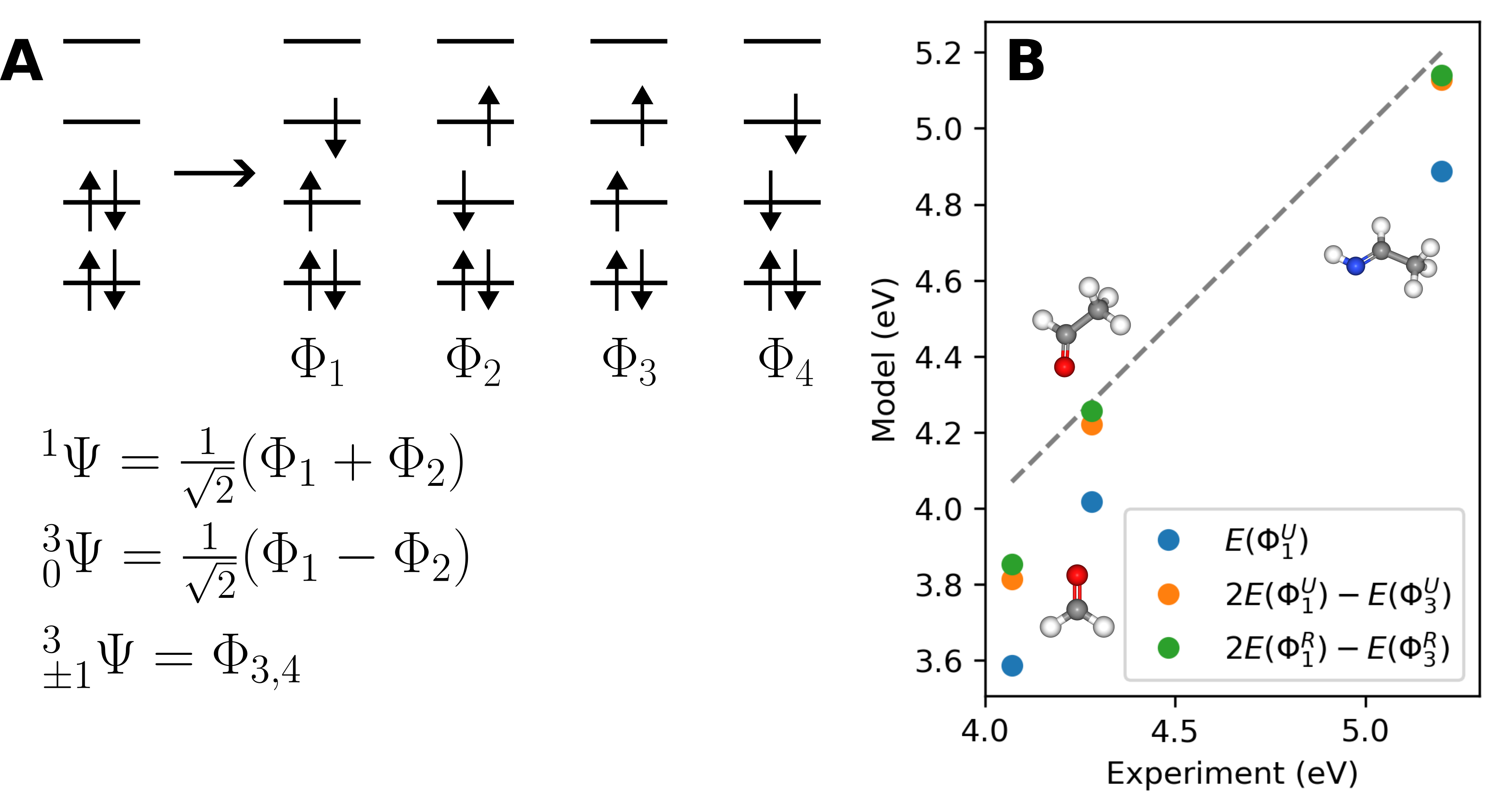}
    \caption{Panel A: Electronic configurations after HOMO to LUMO excitation from ground state reference. The configuration state functions $^1\Psi$ and $^3_{M_s}\Psi$ are the excited singlet and triplet eigenfunctions (with magnetic quantum number $M_s$) and $\Phi_i$ denotes Slater determinants for specific excitations. Panel B: Excitation energies of the lowest excited singlet state of three organic molecules obtained with different approximations (described in text) vs their experimental value~\cite{Robin1985HigherExcitedStatesVol3}. Color coding for molecules: white-hydrogen, gray-carbon, blue-nitrogen, red-oxygen.}
    \label{fig:roks_scheme}
\end{figure}

Additionally, a fundamental conceptual limitation of $\Delta$-SCF approaches is that excited states generally cannot be described by a single Slater determinant.
For instance, the lowest excited singlet state, $^1\Psi$, of a closed shell system must be expressed as a linear combination of two singly excited configurations in order to yield the correct total spin expectation value, $\mel{^1\Psi}{\hat{S}^2}{^1\Psi} = 0$ (see Figure~\ref{fig:roks_scheme}A).
If only one configuration is used, the excitation energy is typically underestimated. This effect is illustrated in Figure~\ref{fig:roks_scheme}B for three organic molecules. When the lowest excited singlet state is approximated solely by the configuration $\Phi_1$ (see Figure~\ref{fig:roks_scheme}A), the resulting excitation energies $E(\Phi_1^U)$ are approximately 200~meV lower than the corresponding experimental values. Nevertheless, accurate singlet energies can be obtained without explicitly evaluating a multideterminantal wave function such as $^1\Psi = \frac{1}{\sqrt{2}} (\Phi_1 + \Phi_2)$. Ziegler \emph{et al.}~\cite{Ziegler1977} demonstrated that the singlet energy can instead be approximated from the energy of the spin-contaminated excited configuration, $E(\Phi_1)$, and that of the corresponding triplet configuration, $E(\Phi_3)$ (see Figure~\ref{fig:roks_scheme}A), according to
\begin{equation}\label{eq:ziegler_sum}
    E_\text{S} = 2 E(\Phi_1) - E(\Phi_3).
\end{equation}
The energies of $\Phi_1$ and $\Phi_3$ can be obtained from two independent unrestricted, i.e. spin-polarized, Kohn-Sham DFT calculations. Inserting these values into Eq.~\eqref{eq:ziegler_sum} yields excitation energies $2E(\Phi_1^U) - E(\Phi_3^U)$ that are in much better agreement with experiment than the uncorrected $E(\Phi_1^U)$ values shown in Figure~\ref{fig:roks_scheme}B. However, this spin-purification procedure is not exact because it is only well founded if the spatial orbitals are identical since this avoids spin contamination. In practice though, the unrestricted calculations yield different orbital sets for $\Phi_1$ and $\Phi_3$, and even within a given configuration the spin-up and spin-down orbitals ($\phi_n^\uparrow \neq \phi_n^\downarrow$) differ. This orbital relaxation leads to residual spin contamination, so that Eq.~\eqref{eq:ziegler_sum} provides only an approximate singlet energy.

A spin-pure singlet state can be obtained by variationally minimizing the full energy expression in Eq.~\eqref{eq:ziegler_sum} under the constraint that the same spatial orbitals are used for both spin channels and for the two electronic configurations $\Phi_1$ and $\Phi_3$. This approach is known as restricted open-shell Kohn–Sham DFT (ROKS)~\cite{Filatov1998,Frank1998,Filatov1999,molecules29184509}.
For the three molecules shown in Figure~\ref{fig:roks_scheme}, the ROKS excitation energies (labeled $2E(\Phi_1^R) - E(\Phi_3^R)$ in Figure~\ref{fig:roks_scheme}) are approximately 30~meV higher than those obtained from two independent unrestricted calculations, corresponding to a slight improvement in agreement with experiment.
In principle, ROKS is a more rigorous approach because it fully removes spin contamination. Moreover, it yields spin-adapted orbitals and densities, which can serve as a consistent starting point for studying exciton dynamics~\cite{PhysRevB.101.100101,C9CP06419B,D3CP00533J,10.1063/5.0288340}.
Despite these advantages, ROKS is not as widely implemented as unrestricted DFT. One likely reason is the increased complexity of the optimization procedure. The orbitals that minimize Eq.~\eqref{eq:ziegler_sum} cannot be obtained by simply solving the standard Kohn–Sham equations, ${\hat{H}^\text{KS} \ket{\phi_n} = \epsilon_n \ket{\phi_n}}$, because the energy functional depends on multiple configurations~\cite{Kowalczyk2013,Hirao1973}.
The correct variational conditions and the associated eigenvalue problem for open-shell systems were originally derived by Roothaan~\cite{Roothaan1960} and later extended to DFT for ground and excited states~\cite{Filatov1998,Frank1998,Filatov1999}. In practice, Eq.~\eqref{eq:ziegler_sum} can be minimized either by direct orbital optimization using, for example, a direct inversion in the iterative subspace (DIIS) solver~\cite{Frank1998}, or by solving the generalized SCF eigenvalue problem associated with the corresponding effective ROKS-Hamiltonian~\cite{Kowalczyk2013,Hirao1973}.

In the following, we present an approach for computing excited singlet-state energies and forces by variationally minimizing Eq.~\eqref{eq:ziegler_sum} using a preconditioned conjugate-gradient algorithm (CG). Our method builds upon the direct minimization scheme for ground-state energies introduced in Ref.~\cite{Kresse1996} and has been implemented in the VASP code~\cite{Kresse1996,PhysRevB.59.1758,PhysRevB.54.11169}. Additionally, we have implemented the optimization procedure via a DIIS solver from Ref~\cite{Frank1998}.
The technical details of the implementation are described in Section~\ref{sec:method}. Finally, we derive analytical expressions for the excited-state forces that are compatible with the projector augmented-wave (PAW) formalism~\cite{PhysRevB.50.17953,PhysRevB.59.1758}. As detailed in Section~\ref{sec:method}, the implementation of ROKS forces leverages most of the existing VASP routines for restricted and unrestricted DFT forces, making the implementation both efficient and highly optimized.
In Section~\ref{sec:results}, we benchmark excited-state energies obtained with our implementation against results from the ROKS implementation in Q-Chem~\cite{shao2015,epifanovsky2021} for eight organic molecules. Furthermore, we apply our method to compute excited-state properties of MgO containing an oxygen vacancy (the $F^0$ center) and compare our results with those from time-dependent DFT (TDDFT) and several high-level electronic structure methods.

\section{Method}\label{sec:method}

\subsection{Energy minimization}
The energy of an excited singlet state is obtained by minimizing the constrained energy functional
\begin{equation}
E = \sum_L c_L E_L - \sum_{m,n} \gamma_{mn} \left( \mel{m}{S}{n} - \delta_{mn} \right),
\end{equation}
where the second term enforces orbital orthonormality during the optimization. Here, $L$ labels the electronic configurations with corresponding coefficients $c_L$ (i.e. $c_1=2$ and $c_2=-1$ in Eq.~\eqref{eq:ziegler_sum}), and $\gamma_{mn}$ are Lagrange multipliers ensuring $\mel{m}{S}{n} = \delta_{mn}$.
In the PAW method~\cite{PhysRevB.50.17953}, the overlap operator $S$ is defined as
\begin{equation}
S = 1 + \sum_{kl} q_{kl} \dyad{\beta_l}{\beta_k},
\end{equation}
where $\beta_k$ are localized projector functions and the coefficients
\begin{equation}
q_{kl} = \int Q_{kl}(\vec{r}) , d\vec{r}
\end{equation}
are obtained from the localized augmentation functions $Q_{kl}$~\cite{PhysRevB.54.11169}.

The minimization is carried out using a preconditioned Fletcher-Reeves conjugate-gradient algorithm \cite{10.1093/comjnl/7.2.149,Kresse1996}. In each iteration $i$, all orbitals $\{\ket{n}\}$ are updated simultaneously according to
\begin{equation}\label{eq:orbital_update}
\ket{n^{i+1}} = \ket{n^i} + \Delta \ket{s_n^i},
\end{equation}
where $\Delta$ denotes the step size and $\ket{s_n^i}$ is the search direction for orbital $\ket{n^i}$ in iteration $i$. The optimization is terminated once the total energy difference between successive iterations, $E[\{\ket{n^{i+1}}\}]-E[\{\ket{n^{i}}\}]$, falls below a predefined threshold.
The step size $\Delta$ is determined by a line search that minimizes the energy
$E[\{ \ket{n^i} + \Delta \ket{s_n^i} \}]$ 
along the search directions ${\ket{s_n^i}}$. The search direction is constructed as
\begin{equation}\label{eq:cg_search}
    \ket{s_n^i} = \ket{p_n^i} + \frac{\braket{p_n^i}{g_n^i}}{\braket{p_n^{i-1}}{g_n^{i-1}}} \ket{s_n^{i-1}},
\end{equation}
where $\ket{g_n} = \frac{\delta E}{\delta \bra{n}}$ is the constrained energy gradient and $\ket{p_n}$ is the preconditioned gradient, whose explicit form will be specified below.

The constrained energy gradient $\ket{g_n}$ is given by
\begin{equation}\label{eq:gradient_gamma_version}
\begin{split}
        \ket{g_n} &= \sum_L c_L \frac{\delta E_L}{\delta \bra{n}} - \sum_m \gamma_{nm} S\ket{m} \\
        &= \sum_{\sigma L} c_L f_n^{\sigma L} H^{\sigma L} \ket{n} - \sum_m \gamma_{nm} S\ket{m},
\end{split}
\end{equation}
where $\sigma$ denotes the spin channel and $f_n^{\sigma L}$ is the occupation number of the orbital $n$ in configuration $L$ and spin channel $\sigma$ (with $f_n^{\sigma L} \in {0,1}$).
The configuration and spin dependent Hamiltonian in atomic units is defined as
\begin{equation}
    H^{\sigma L} = \hat{T} + \hat{V}_\text{ext} + \int d\vec{r}' \frac{\rho_L(\vec{r}')}{|\vec{r}-\vec{r}'|} + \upsilon^{\sigma}_\text{xc},
\end{equation}
where $\hat{T}$ is the kinetic energy operator and $\hat{V}_\text{ext}$ is the external potential. The third term is the Hartree potential generated by the density $\rho_L$ of configuration $L$, and $\upsilon^{\sigma}_\text{xc}$ is the spin-dependent exchange–correlation potential for that density.

The Lagrange multipliers $\gamma_{nm}$ are determined from the stationarity condition at the energy minimum,
\begin{equation}\label{eq:stationary_cond}
    dE = \sum_n \braket{dn}{g_n} + \braket{g_n}{dn} = 0,
\end{equation}
where $\ket{d n}$ denotes an arbitrary orbital variation.
As shown in more detail in the supplementary information (SI) section S1, this leads to the stationarity condition
\begin{equation}\label{eq:hermitian_condition}
    \braket{g_m}{n} + \braket{m}{g_n} = 0
\end{equation}
for non-unitary orbital variations.
Substitution of $\ket{g_n}$ in  Eq.~\eqref{eq:hermitian_condition} by its definition in Eq.~\eqref{eq:gradient_gamma_version} yields
\begin{equation}\label{eq:gamma_def}
    \gamma_{nm} = \frac{1}{2} \sum_{L,\sigma} (f_{n,L}^\sigma + f_{m,L}^\sigma) c_L \mel{m}{H_L^\sigma}{n}.
\end{equation}
This definition of $\gamma$ was already derived in Ref~\cite{Kresse1996} for a single spin-channel and a single electronic configuration $L$ with $c_L=1$. However, we note that the stationarity condition Eq.~\eqref{eq:hermitian_condition} (Eq.~112 in Ref~\cite{Kresse1996}) erroneously  contained $\braket{n}{g_m} $ instead of $\braket{g_m}{n}$. Furthermore, it was stated that this condition is necessary if unitary transformations are allowed, while it is necessary if \emph{non}-unitary transformations are allowed (see SI section S1).
Nevertheless, the final definition of $\gamma$ reported in \cite{Kresse1996} agrees with the definition derived in the present study.

Using this definition of $\gamma$, the gradient can be written as
\begin{equation}\label{eq:gradient}
\begin{split}
    \ket{g_n} &= \sum_{\sigma,L} c_L \Bigg[
    \underbrace{f_n^{\sigma L} \left(1 - \sum_m S\dyad{m} \right) H^{\sigma L} \ket{n}}_{:= \ket{g_{n\mathrm{I}}}} \\
    &\quad + \underbrace{\sum_m \frac{1}{2} (f_n^{\sigma L} - f_m^{\sigma L}) H_{mn}^{\sigma L} S\ket{m}}_{:= \ket{g_{n\mathrm{II}}}}
    \Bigg].
\end{split}
\end{equation}
The first contribution, $\ket{g_{n\mathrm{I}}}$, represents the component of the gradient that lies outside the subspace spanned by the current orbitals $\{\ket{m}\}$. In contrast, $\ket{g_{n\mathrm{II}}}$ describes, to first order, the energy change associated with unitary rotations within this subspace. 
For a closed-shell ground state (e.g., in an insulator) and if only occupied orbitals are included in $\{\ket{m}\}$, one has $f_n^{\sigma L} = f_m^{\sigma L}$ for all relevant orbitals, and $\ket{g_{n\mathrm{II}}}$ vanishes.
This reflects the invariance of the total energy under unitary transformations of the occupied subspace. However, for open-shell systems or when unoccupied orbitals are included in the set $\{\ket{m}\}$, the occupations differ and $\ket{g_{n\mathrm{II}}}$ gives a finite contribution to the energy variation.

Furthermore, by defining the orbital-dependent operators
\begin{equation}\label{eq:shell_operator}
    F^n = \sum_{\sigma L} c_L f_n^{\sigma L} H^{\sigma L},
\end{equation}
we obtain for $\ket{g_{nI}} = 0$
\begin{equation}\label{eq:stat1}
    F^n \ket{n} = \sum_m S\ket{m} \mel{m}{F^n}{n}.
\end{equation}
Requiring the subspace gradient to be zero, i.e. $\ket{g_{nII}} = 0$ and multiplication by $\bra{m}$ leads to
\begin{equation}\label{eq:stat2}
    \mel{m}{F^n - F^m}{n} = 0.
\end{equation}
The equations~\eqref{eq:stat1} and \eqref{eq:stat2} are identical to the stationarity conditions of the generalized SCF operator derived in Ref~\cite{Hirao1973}.

The decomposition of the gradient into out-of-subspace and in-subspace components enables more effective preconditioning of the minimization procedure~\cite{Kresse1996}.
Analogously to Ref.~\cite{Kresse1996}, we construct a preconditioned gradient for the out-of-subspace component, $\ket{p_{n\mathrm{I}}}$, as
\begin{equation}\label{eq:prec_grad_I}
    \ket{p_{nI}} =K \ket{g_{nI}} - \sum_m \ket{m}\mel{m}{SK}{g_{nI}}
\end{equation}
with the preconditioner $K$ defined as~\cite{PhysRevB.40.12255}
\begin{equation}\label{eq:prec_k}
    K = -\sum_{\vec{q}} \frac{2 \dyad{q} }{ 10 } \frac{27 + 18 x + 12 x^2 + 8 x^3}{27 + 18 x + 12 x^2 + 8 x^3 + 16 x^4},
\end{equation}
where
\begin{equation}
    x = \frac{E^\text{kin}(\vec{q})}{10 }
\end{equation}
and $E^\text{kin}(\vec{q})$ being the kinetic energy of the plane wave $\vec{q}$ in Rydberg atomic units.
The second term in Eq.~\eqref{eq:prec_grad_I} removes components of $K\ket{g_{n\mathrm{I}}}$ that lie within the subspace spanned by the current orbitals $\{\ket{m}\}$. This guaranties that the preconditioned search direction does not violate the orthogonality constraints to first order.

Instead of following the in-subspace gradient component $\ket{g_{n\mathrm{II}}}$, the energy within the subspace spanned by the current orbitals $\{\ket{m}\}$ can be more efficiently minimized using iterative matrix diagonalization techniques.
For a closed-shell ground state, the energy is minimized when the orbitals are eigenvectors of the Hamiltonian represented in the basis $\{\ket{m}\}$. These optimized orbitals ${\ket{n'}}$ can be obtained by a unitary transformation,
\begin{equation}\label{eq:inspace_rot}
\ket{n'} = \sum_m U_{mn} \ket{m^i},
\end{equation}
where $U$ rotates the current orbitals ${\ket{m^i}}$ into the eigenvectors of the Hamiltonian. In Ref.~\cite{Kresse1996}, this rotation is performed approximately at each conjugate-gradient step using a scheme related to L\"owdin perturbation theory and the Jacobi diagonalization algorithm~\cite{Jacobi}.

We adopt the same strategy here, but with an important modification. In contrast to the ground state, the excited singlet energy defined by Eq.~\eqref{eq:ziegler_sum} is not minimized by the eigenvectors of the Kohn–Sham Hamiltonian, nor by those of the weighted operator $\sum_{L,\sigma} c_L H^{\sigma L}$. Instead, the minimizing orbitals ${\ket{n^{\mathrm{min}}}}$ are eigenvectors of the matrix~\cite{Kowalczyk2013,Hirao1973}
\begin{equation}\label{eq:r_matrix_singlet}
    R = \mqty( F^c & F^c - F^a & F^c - F^b & F^c \\
               F^c - F^a & F^a & F^a - F^b & F^a \\ 
               F^c - F^b & F^a - F^b & F^b & F^b \\
               F^c & F^a & F^b & F^c
               ),
\end{equation}
where the shell-specific operators are defined in Eq.~\eqref{eq:shell_operator}.
Each shell groups orbitals sharing the same occupation pattern ${f_n^{\sigma L}}$ in all configurations. For an excited singlet state, these shells comprise a closed shell ($F^c$), two open shells ($F^a$ and $F^b$), and an empty shell that contains the unoccupied orbitals. The diagonal blocks of $R$ correspond to couplings within a shell, whereas the off-diagonal blocks describe couplings between different shells~\cite{Hirao1973,DanielEdwards1987,Kowalczyk2013}. Excited high-spin states (e.g. a triplet state) can be modeled by considering only one open shell in the definition of $R$.

To render the energy stationary, we seek a unitary transformation that rotates the orbitals into the eigenvectors of $R$. We therefore define the unitary matrix
\begin{equation}\label{eq:prec_grad_IIa}
    U = e^{\Delta A^i},
\end{equation}
where $\Delta$ is the step size (cf.~Eq.~\eqref{eq:orbital_update}) and $A^i$ is an anti-Hermitian matrix defined in iteration $i$ with
\begin{equation}\label{eq:prec_grad_IIb}
    A^i_{mn} = (1-\delta_{nm}) \arctan(2 x)/2 
\end{equation}
and
\begin{equation}\label{eq:prec_grad_IIc}
    x = \frac{R_{mn}}{\sum_{\sigma L} c_L (H^{\sigma L}_{mm}-H^{\sigma L}_{nn})}.
\end{equation}
The orbitals are iteratively transformed into the eigenvectors of the matrix defined in Eq.~\eqref{eq:r_matrix_singlet} by applying this matrix $U$ once at each conjugate gradient step.

The orbital update is performed using the conjugate-gradient algorithm already implemented in VASP. In this scheme, the orbitals are updated in two successive steps.
First, the out-of-subspace update is carried out according to Eq.~\eqref{eq:orbital_update}, using the (preconditioned) gradients $\ket{g_{n\mathrm{I}}}$ and $\ket{p_{n\mathrm{I}}}$ (see Eqs.~\eqref{eq:gradient} and \eqref{eq:prec_grad_I}).
Second, the in-subspace optimization is performed via the unitary transformation defined in Eq.~\eqref{eq:inspace_rot}.
The unitary matrix is constructed as
\begin{equation}
    U = e^{\Delta (A^i+\gamma A^{i-1})},
\end{equation}
where $\gamma = \frac{\braket{p_n^i}{g_n^i}}{\braket{p_n^{i-1}}{g_n^{i-1}}}$. Here, $\ket{g_n}$ and $\ket{p_n}$ denote the full gradients, i.e., the sums of the in-subspace and out-of-subspace (preconditioned) contributions. The matrix $A^i$ is defined according to Eqs.~\eqref{eq:prec_grad_IIb} and \eqref{eq:prec_grad_IIc}, while $A^{i-1}$ denotes the corresponding matrix from the previous iteration.

By defining $\theta_{mn} = \mel{m}{F^n}{n}$, the gradient in Eq.~\eqref{eq:gradient} can also be expressed as
\begin{equation}
    \ket{g_n} = F_n \ket{n} - \sum_m \theta_{mn} S\ket{m} + \frac{1}{2} \sum_m (\theta_{mn} - \theta_{nm}^*) S\ket{m}.
\end{equation}
At $\ket{g_n} = 0$, this yields
\begin{equation}\label{eq:residual_frank}
    F^n \ket{n} = \sum_m S\ket{m} [\theta_{mn} + \lambda_{ji} \sum_m (\theta_{mn} - \theta_{nm}^*) ]
\end{equation}
with $\lambda_{ji} = \frac{1}{2}$. This is identical to the variational conditions for the restricted open shell singlet derived in Refs~\cite{Frank1998,Grimm2003}. In Refs~\cite{Frank1998,Grimm2003}, a DIIS method~\cite{Hutter1994} was used to minimize the gradient in Eq.~\eqref{eq:residual_frank}. We have implemented this DIIS solver as well as described in Ref~\cite{Hutter1994} in a development version of VASP. The implementation details will be reported elsewhere. The advantage of the DIIS algorithm is that it can converge onto saddle points while the conjugate gradient algorithm can only reliably converge towards a minimum. Hence, the DIIS implementation can be used to model excited states beyond the lowest excited singlet state that can correspond to saddle points with respect to variations in the orbitals \cite{Schmerwitz2023}.

Nevertheless, complications can arise because the ROKS formalism does not explicitly enforce orthogonality between the excited singlet state and the ground state \cite{Carbo1978}. Consequently, the coupling between the two open-shell orbitals can drive the system toward an unphysical "triplet-like" state during the minimization of Eq.~\eqref{eq:ziegler_sum}. In this regime, the excited singlet energy becomes nearly degenerate with the triplet state, and the resulting wave function can exhibit significant overlap with the ground state \cite{Kowalczyk2013}. 
To steer the system away from this triplet-like state and toward an energetically higher-lying state that better represents the excited singlet, Kowalczyk \emph{et al.}~\cite{Kowalczyk2013} proposed suppressing the coupling between the two open-shell orbitals during minimization. We have adapted this strategy for our conjugate gradient algorithm by setting the matrix element $F_a-F_b$ in Eq.~\eqref{eq:r_matrix_singlet} to zero. For minimization via the DIIS solver, instead we choose different coupling coefficients $\lambda_{ab} = 1/2$ and $ \lambda_{ba} = -1/2$ in Eq.~\eqref{eq:residual_frank} for the two open shell orbitals as suggested in \cite{Grimm2003,FRIEDRICHS200817}.

\subsection{Forces}
The derivative of the total energy with respect to a nuclear coordinate $R_{I\nu}$ where $I$ is the atom index and $\nu$ the coordinate, i.e., the negative force $\frac{dE}{dR_{I\nu}} = -F_{I\nu}$, is given by
\begin{equation}\label{eq:force_general}
    \frac{dE}{dR_{I\nu}} = \sum_L c_L \frac{d E_L}{d R_{I\nu}} - \sum_{m,n \in \text{occ}} \frac{d}{dR_{I\nu}} \gamma_{mn} \left(\mel{m}{S}{n} - \delta_{mn}\right).
\end{equation}
The summation in the second term can be restricted to occupied orbitals, since only these contribute to the energy and its derivative. Because the energy is stationary with respect to orbital variations, the forces can be evaluated using a generalized Hellmann–Feynman theorem (see Ref.~\cite{Goedecker1992} for a detailed derivation).

This yields
\begin{equation}\label{eq:force_hf}
\begin{split}
            -F_{I\nu} = \frac{dE}{dR_{I\nu}} &= \sum_{L\sigma n} c_L f_n^{\sigma L} \mel{n}{\frac{\partial H^{\sigma L}}{\partial R_{I\nu}}}{n} \\
            &- \sum_{m,n \in \text{occ}} \gamma_{mn} \mel{m}{\frac{\partial S}{\partial R_{I\nu}}}{n} .
\end{split}
\end{equation}
The second term can be efficiently calculated by transforming the complex-conjugate $\bar{\gamma}$ to a basis $\ket{n'}$ in which the matrix $\gamma$ is diagonal, as shown in the SI section S2. In this basis, Eq.~\eqref{eq:force_hf} becomes
\begin{equation}\label{eq:force_hf2}
\begin{split}
            -F_{I\nu} = \frac{dE}{dR_{I\nu}} &= \sum_{L\sigma n} c_L f_n^{\sigma L} \mel{n}{\frac{\partial H^{\sigma L}}{\partial R_{I\nu}}}{n} \\
            &-  \sum_{n \in \text{occ}}  \gamma'_{nn} \mel{n}{ U^\dagger\frac{\partial S}{\partial R_{I\nu}}U}{n},
\end{split}
\end{equation}
where $\gamma' = U^\dagger \bar{\gamma} U$ and $U$ is the unitary transformation making $\bar{\gamma}$ (and also $\gamma$) diagonal.

The resulting expression, Eq.~\eqref{eq:force_hf2}, closely resembles the force formula for restricted and unrestricted DFT within the PAW formalism, where $\gamma$ is diagonal for variationally optimized orbitals~\cite{PhysRevB.59.1758}. Consequently, ROKS forces can be implemented with only minor modifications to the existing force routines in VASP.

\section{Computational Details}\label{sec:comp_det}
The ROKS calculations were performed with a development version of VASP. 
The interaction between the valence electrons and the ionic cores was accounted for by the projector augmented wave method \cite{PhysRevB.50.17953} with the PAW potentials described in Table~\ref{tab:paw-potentials}.

\begin{table}[htbp]
\centering
\caption{The PAW potential used in the present work. Valence specifies the orbitals treated as valence orbitals and label the respective PAW potential in the VASP database. The radial cutoffs for each angular momentum quantum number are specified as $n \times r_{\mathrm{cut}}$ with $n$ denoting the number of projectors and $r_{\mathrm{cut}}$ being the respective radial cutoff in atomic units. The local potential corresponds to the all-electron potential, which is replaced by a local pseudopotential below the radius $r_{\mathrm{core}}$ (a.u.).}
\label{tab:paw-potentials}
\begin{tabular}{ccccccc}
\hline\hline
atom & valence & label & $s$ & $p$ & $d$ & $r_{\mathrm{core}}$ \\
\hline
H  & $1s^1$& PBE & $2 \times 1.1$ & $1 \times 1.1$ & -- & 1.1 \\
C  & $2s^2 2p^2$ & PBE & $2 \times 1.2$ & $2 \times 1.5$ & $1 \times 1.5$ & 1.5 \\
N  & $2s^2 2p^3$ & PBE & $2 \times 1.2$ & $2 \times 1.5$ & $1 \times 1.5$ & 1.5 \\
O & $2s^2 2p^4$ & PBE & $2 \times 1.2$ & $2 \times 1.52$ & $1 \times 1.5$ & 1.52 \\
Mg  & $3s^2$ & PBE & $2 \times 2.0$ & $2 \times 2.0$ & $1 \times 2.0$ & 2.0 \\
\hline\hline
\end{tabular}
\end{table}
For the isolated organic molecules, a broken-symmetry orthorhombic supercell was utilized with dimensions of 15.0~\AA $\times$ 15.3~\AA $\times$ 15.5~\AA \ to minimize spurious interactions between periodic images and the system was only sampled at the $\Gamma$-point. A plane-wave cutoff of $750$~eV was used. The convergence criteria for the electronic energy and the interatomic forces were set to $10^{-6}$~eV and $0.01$~eV/\AA, respectively. Using these convergence criteria, the geometries were optimized in the ground state. The vertical excitation energies were then calculated using the orbitals from the ground state calculation as initial guesses for the excited state orbitals. For the excited state calculations of acetaldimine, ethene, and formamide using the PBE0 functional, we used the ground state orbitals from the LDA calculations rather than the PBE0 ground state orbitals. In all other cases, the excited state calculations used ground state orbitals obtained with the same functional.

All Q-Chem calculations were single-point calculations carried out using Q-Chem version 6.3.0~\cite{shao2015,epifanovsky2021,Kowalczyk2013}. The molecular geometries were taken from prior optimizations performed with VASP, as described above, and were used without further modification. DFT and ROKS calculations were subsequently performed using the def2-QZVP basis set throughout. In the ROKS calculations, we used the squared-gradient minimization algorithm~\cite{doi:10.1021/acs.jctc.9b01127}.

The MgO system containing a neutral oxygen vacancy was modeled using a cubic $3 \times 3 \times 3$ supercell with 215 atoms. The defect was created by removing a single oxygen atom from the bulk structure. 
The lattice constant for this defective cell was maintained at a fixed value of 4.19\AA, which is the lattice constant of MgO obtained with the PBE0 functional with 36\% Fock-exchange \cite{PhysRevB.89.195112}.
In all calculation only the $\Gamma$-point was sampled. During the geometry optimization, the cell volume and shape were kept constant while the internal atomic positions were allowed to relax. The threshold for the interatomic forces was always set to $0.01$~eV/\AA. The ground state was optimized either with PREC set to normal or to accurate and the energy convergence was set to $10^{-5}$~eV. The excited state where an electron is moved to the conduction band minimum  (CBM) was optimized with the conjugate gradient algorithm, PREC set to normal and the energy convergence set to $10^{-5}$~eV. The $T_{1u}$-states were optimized with the DIIS algorithm with PREC set to accurate, the energy convergence was set to $10^{-6}$~eV and no symmetry constraints were enforced during the geometry optimization of the $T_{1u}$-states.

The calculations were performed with the PBE and the PBE0 functional. In the PBE0 calculations, the amount of global exact Fock exchange was set to $\alpha = 0.36$, which is equal to the inverse dielectric constant of MgO \cite{PhysRevB.89.195112}. In this manuscript, we refer to this functional as dielectric dependent hybrid (DDH) functional.

\section{Results}\label{sec:results}

\subsection{Benchmarking excitation energies in organic molecules}\label{sec:resA}
To validate our implementation, we compared excitation energies to the lowest excited singlet state of eight organic molecules, calculated using the CG and DIIS implementations in VASP and ROKS as implemented in Q-Chem~\cite{Kowalczyk2013,Hait2020}.
As one would expect, the CG and the DIIS implementation yielded identical excitation energies $(< 5\cdot 10^{-5})$ except for cytosine; the reasons for the deviations in this case will be discussed below. However, the CG implementation converged in significantly less iterations. For example, converging the singlet state of acetaldehyde with the PBE0 functional and a convergence threshold of $10^{-6}$ required only 18 iterations with the CG algorithm but 64 iterations with DIIS.
The reason is most likely the better preconditioning of the CG approach based on the splitting of the gradient into an out-of-subspace and in-subspace part. In contrast, the DIIS solver uses only the preconditioner $K$ defined in Eq.~\eqref{eq:prec_k}.

\begin{figure}
    \centering
    \includegraphics[width=1.0\linewidth]{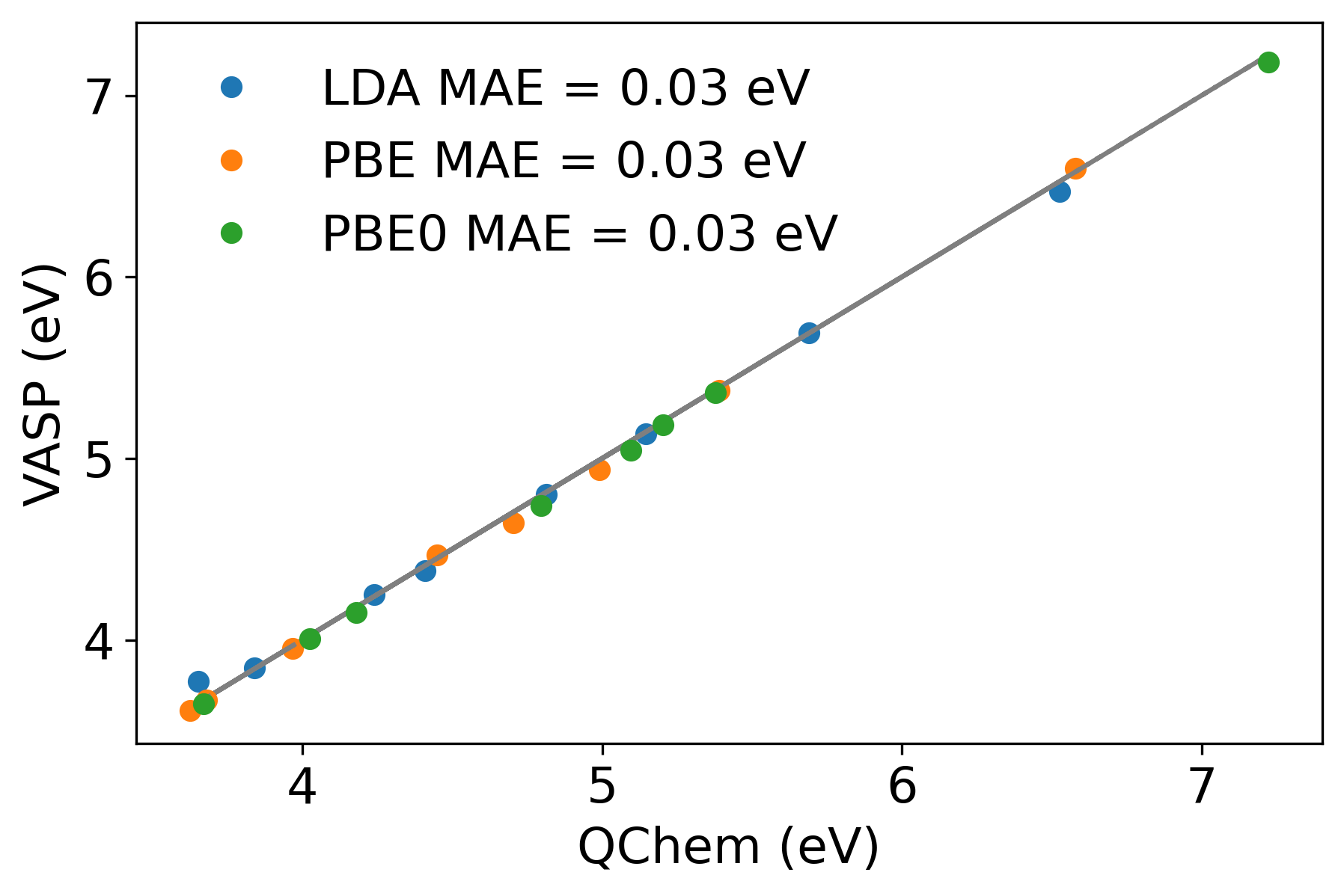}
    \caption{Excitation energies to the lowest lying singlet state for eight small organic molecules computed with ROKS as implemented in VASP (this work) and Q-Chem for different DFT functionals.}
    \label{fig:vasp_vs_qchem}
\end{figure}

Furthermore, the overall agreement between the VASP and Q-Chem implementations is very good. The mean absolute deviations amount to 30, 25, and 31 meV for LDA, PBE, and PBE0 functionals, respectively (see Figure~\ref{fig:vasp_vs_qchem} and Table~\ref{tab:vasp_vs_qchem}). The remaining differences are most likely due to distinct numerical approximations. For instance, VASP uses a plane-wave basis set and PAW potentials, whereas Q-Chem employs Gaussian-type orbitals and treats all electrons explicitly.
The impact of such methodological differences has previously been analyzed for atomization energies~\cite{10.1063/1.1926272,doi:10.1021/acs.jctc.3c00089}. In these studies, mean deviations between VASP and all-electron Gaussian-based codes of 20 up to 120~meV were reported, which is comparable to the differences observed here for excitation energies.

For the $\pi \rightarrow \pi^*$ excitations in ethene, butadiene, and cytosine, the excitation energies can be underestimated due to a collapse to a triplet-like state as described in the Method section.
For example, during the minimization of Eq.~\eqref{eq:ziegler_sum} for ethene with the CG algorithm the two open shell orbitals localized on different parts of the molecule. Additionally, the energy converged towards the energy of the triplet state as has been observed in Refs~\cite{Grimm2003,FRIEDRICHS200817}. 
To avoid this behavior when using the CG algorithm, we suppressed the coupling between the two open shells as suggested in Ref~\cite{Kowalczyk2013}. However, this procedure does not guaranty that the resulting energy is stationary with respect to orbital variations because the rotations between the open shells are neglected. Therefore, we verified that the obtained solution was stationary by evaluating the open shell coupling element in Eq.~\eqref{eq:r_matrix_singlet} after energy convergence. The coupling element should be zero according to Eq.~\eqref{eq:stat2}. Indeed, the values were quite small ($\leq 10^{-4}$~eV) for ethene and butadiene indicating that the obtained solution was indeed stationary. 
If the DIIS algorithm was employed, we obtained the same excitation energies as with the CG method for ethene and butadiene by setting $\lambda_{ab} = \frac{1}{2}$ and $\lambda_{ba} = -\frac{1}{2}$ in Eq.~\eqref{eq:residual_frank} as suggested in Ref~\cite{Grimm2003}.

However, modelling of the excited singlet state in cytosine turned out to be more challenging.
If the open shell coupling was suppressed when using the CG algorithm, the coupling did not decay to zero during the optimization but was still in the order of 0.1~eV after the other degrees of freedom where optimized. 

The DIIS optimization of cytosine behaved differently with the different functionals.
For the PBE0 functional, we obtained an excitation energy of 4.149~eV, close to the value obtained from the CG algorithm if the coupling was suppressed. However, the open shell coupling element ($\approx 10^{-3}$~eV) was significantly smaller than the one obtained from the CG algorithm, indicating that the DIIS solution corresponded to a stationary solution. 
The LDA and PBE functionals yielded excitation energies of 3.663 and 3.507~eV. These values are identical to the energies obtained with the CG algorithm if the coupling between the two open shells is \emph{not} suppressed. This indicates that for cytosine, the collapse to a lower lying triplet-like state cannot be avoided for the LDA and PBE functional even by setting $\lambda_{ab} = \frac{1}{2}$ and $\lambda_{ba} = -\frac{1}{2}$. 

However, it was possible to obtain the singlet state by combining the CG and DIIS algorithms. First, the state was optimized with the CG algorithm and suppressed open shell coupling. The obtained orbitals were then further refined with the DIIS algorithm. This yielded stationary states (open shell couplings $\approx 10^{-3}$~eV) with excitation energies of 3.772 and 3.669~eV for the LDA and PBE functionals, respectively.

The LDA Q-Chem calculation yielded an excitation energy of 3.654~eV, close to the value from the VASP CG calculation if the coupling was not suppressed (3.663~eV).
This indicates a collapse to the triplet-like state. Indeed, the wave function of the excited state from the Q-Chem LDA calculation had a large overlap with the ground state wave function $(\braket{\Psi_{\text{S0}}}{\Psi_{\text{S1}}} \approx 0.49)$ confirming the collapse.

Overall, these results demonstrate that our implementation can reliably predict the lowest excited singlet state if it has a symmetry different from that of the ground state. If the excited state and the ground state have the same symmetry, the coupling between the open shells can lead to an underestimation of the excitation energies. In such a case the singlet state can be obtained either by setting the open shell coupling element to zero if the CG algorithm is used or by choosing different couplings $\lambda_{ab} \neq \lambda_{ba}$ between the open shells if the DIIS method is used. Sometimes, a combination of the two approaches is necessary.

\begin{table*}[htbp]
\centering
\caption{Comparison of excitation energies to the lowest lying excited singlet states (in eV) obtained from VASP (planewave cutoff 750 eV) and Q-Chem (def2-QZVP basis) with ROKS for different organic molecules using three density functionals. The $\Delta$ column contains the differences between the VASP and Q-Chem values. The VASP results for cytosine are obtained with the DIIS implementation for PBE0 and by a combination of the CG and DIIS algorithm for LDA and PBE, as explained in the text. The other results are obtained with the VASP CG algorithm.}
\begin{tabular}{cccccccccc}
\hline\hline
\multirow{2}{*}{\textbf{Compound}} &
\multicolumn{3}{c}{\textbf{LDA}} &
\multicolumn{3}{c}{\textbf{PBE}} &
\multicolumn{3}{c}{\textbf{PBE0}} \\
 & VASP & Q-Chem & $\Delta$ & VASP & Q-Chem & $\Delta$ & VASP & Q-Chem & $\Delta$ \\
\hline
acetaldehyde & 4.248 & 4.239 & $0.009$ & 3.954 & 3.969 & -0.014 & 4.006 & 4.024 & $-0.018$ \\
acetaldimine & 5.136 & 5.145 & $-0.009$ & 4.939 & 4.992 & $-0.053$ & 5.045 & 5.097 & $-0.052$ \\
butadiene & 4.383 & 4.410 & $-0.027$ & 4.467 & 4.448 & 0.019 & 5.184 & 5.203 & $-0.019$ \\
cytosine & 3.772 & 3.654 & 0.118 & 3.669 & 3.681 & $-0.012$ & 4.149 & 4.180 & $-0.032$ \\
ethene & 6.472 & 6.525 & $-0.053$ & 6.597 & 6.578 & 0.018 & 7.185 & 7.222 & $-0.037$ \\
formaldehyde & 3.846 & 3.840 & 0.006 & 3.609 & 3.626 & $-0.017$ & 3.649 & 3.669 & $-0.020$ \\
formaldimine & 4.802 & 4.815 & $-0.013$ & 4.647 & 4.704 & $-0.057$ & 4.742 & 4.797 & $-0.055$ \\
formamide & 5.692 & 5.689 & 0.004 & 5.375 & 5.390 & $-0.014$ & 5.362 & 5.378 & $-0.016$ \\
\hline
MAE & & & 0.030 & & & 0.025 & & & 0.031 \\
\hline\hline
\end{tabular}
\label{tab:vasp_vs_qchem}
\end{table*}

Having established the accuracy of the excitation energies, we finally validated the analytical force implementation. To this end, we compared the forces obtained from Eq.~\eqref{eq:force_hf} for acetaldehyde in the ground state equilibrium geomtry with numerical forces computed via finite differences.
The agreement is excellent, with mean absolute deviations between the analytical and finite difference forces of 0.1, 0.1 and 0.3~meV/Ang and relative deviations of 0.7\%, 4.2\% and 1.1\% for LDA, PBE and PBE0, respectively.

\subsection{Excitation energies of the MgO $F^0$-center}

\begin{figure}
\centering
\includegraphics[width=1.0\linewidth]{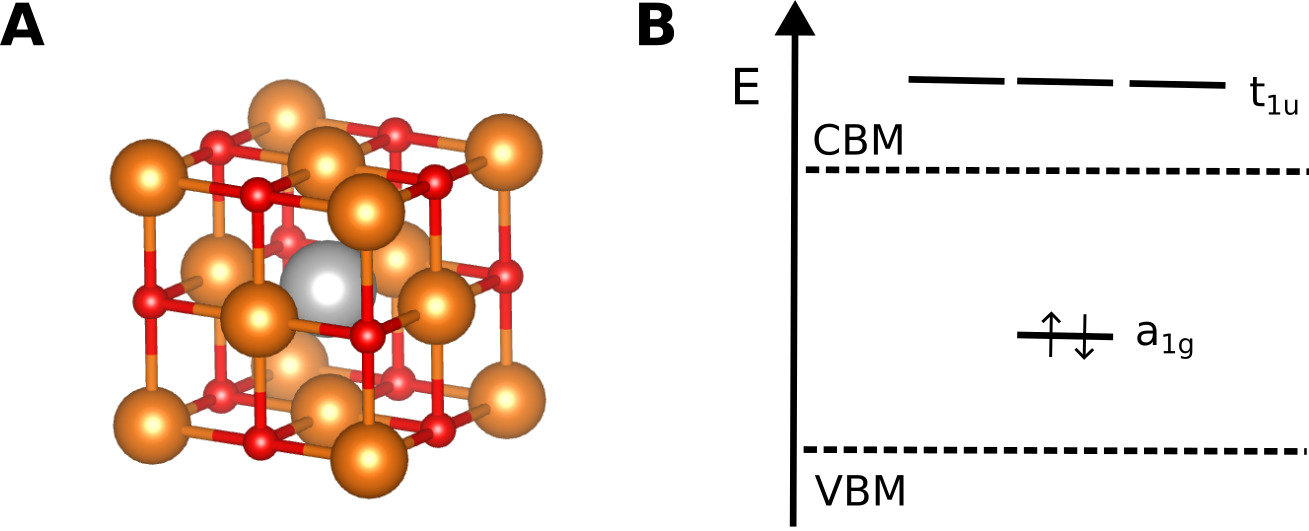}
\caption{A: Structure of MgO with oxygen vacancy $V_\text{O}$ (in gray). B: The orbitals at the $\Gamma$-point close to the valence band maximum (VBM) and conduction band minimum (CBM) of the structure in A.}
\label{fig:MgO_structure}
\end{figure}

After validating our implementation for small molecules, we applied the method to calculate excitation energies of magnesium oxide containing a neutral oxygen vacancy, i.e., the $F^0$-center. In its ground state, this defect exhibits octahedral symmetry (see Figure~\ref{fig:MgO_structure}A). A schematic band structure at the $\Gamma$-point, focusing on states near the Fermi level, is shown in Figure~\ref{fig:MgO_structure}B. The oxygen vacancy introduces a localized $a_{1g}$ defect state within the band gap, which is occupied by two electrons. In addition, three degenerate unoccupied $t_{1u}$ defect states appear somewhat above the conduction band minimum and are likewise localized around the vacancy.

Experimentally, optical absorption has been observed at 5.03~eV, while emission peaks have been reported at approximately 2.4 and $3.1-3.2$~eV. A recent linear-response TDDFT study investigated the lowest excitation to the conduction band minimum as well as the lowest singlet and triplet excitations of $T_{1u}$ symmetry~\cite{galli_tddft}. In particular, that work reported vertical excitation energies, Franck--Condon shifts, emission energies, and mass-weighted displacements between the ground- and excited-state equilibrium geometries. We computed the same set of properties using our ROKS implementation. The results are summarized in Table~\ref{tab:MgO} together with the TDDFT reference data~\cite{galli_tddft}. All results are compared for a $3\times3\times3$ supercell of the conventional cell containing 215 atoms. The TDDFT results from Ref~\cite{galli_tddft} at this cell size are obtained from https://paperstack.uchicago.edu.

\begin{table*}[ht]
\centering
\caption{Excited state properties of MgO with a neutral oxygen vacancy for excitations to the $^1T_{1u}$, $^3T_{1u}$-states and the conduction band minimum (CBM). Vertical absorption energies, $E_\text{abs}$, emission energies, $E_\text{emi}$, Franck-Condon shifts in the ground and excited state, $E_{\mathrm{FC,GS}}$ and $E_{\mathrm{FC,ES}}$, and mass weighted displacements $\Delta Q$ are calculated with ROKS (this work) and TDDFT (from Ref~\cite{galli_tddft}) using PBE and the dielectric dependent hybrid functional (DDH = PBE0 with 36\% Fock exchange).}
\setlength{\tabcolsep}{6pt}
\renewcommand{\arraystretch}{1.2}
\begin{tabular}{llccccc}
\toprule
\multirow{2}{*}{states} & \multirow{2}{*}{method} & $E_{\mathrm{abs}}$ & $E_{\mathrm{emi}}$ & $E_{\mathrm{FC,GS}}$ & $E_{\mathrm{FC,ES}}$ & $\Delta Q$ \\
& & (eV) & (eV) & (eV) & (eV) & (amu$^{1/2}$\AA)\\
\midrule
\multirow{2}{*}{$^{1}T_{1u}$}
& ROKS-PBE & 4.235 & 3.476 & 0.382 & 0.377 & 1.136 \\
& ROKS-DDH & 4.928 & 4.279 & 0.317 & 0.332 & 1.222 \\
& TDDFT-PBE & 3.857 & 2.910 & 0.410 & 0.537 & 1.694 \\
& TDDFT-DDH & 5.343 & 4.336 & 0.607 & 0.400 & 1.160 \\
\midrule
\multirow{2}{*}{$^{3}T_{1u}$}
& ROKS-PBE & 3.561 & 2.895 & 0.336 & 0.330 & 1.445 \\
& ROKS-DDH & 3.812 & 2.949 & 0.429 & 0.434 & 1.532 \\
& TDDFT-PBE & 3.289 & 2.668 & 0.263 & 0.358 & 2.224 \\
& TDDFT-DDH & 3.793 & 2.943 & 0.419 & 0.431 & 1.573 \\
\midrule
\multirow{4}{*}{CBM}
& ROKS-PBE & 3.036 & 1.588 & 0.690 & 0.758 & 1.257 \\
& ROKS-DDH & 3.969 & 2.129 & 0.905 & 0.935 & 1.342 \\
& TDDFT-PBE & 2.345 & 1.855 & 0.182 & 0.308 & 0.659 \\
& TDDFT-DDH & 4.151 & 2.776 & 0.643 & 0.732 & 1.092 \\
\bottomrule
\end{tabular}
\label{tab:MgO}
\end{table*}

Using our ROKS implementation, we modeled all three excitations considered in Ref.~\cite{galli_tddft}. The triplet $^3T_{1u}$ state was obtained by promoting one electron from the localized $a_{1g}$ defect orbital to one of the unoccupied $t_{1u}$ orbitals with a spin flip. The corresponding singlet $^1T_{1u}$ state was described from the mixed-spin configuration and the associated triplet configuration generated by the $a_{1g}\rightarrow t_{1u}$ excitation. The singlet excitation to the CBM was treated analogously, with the singly occupied orbitals corresponding to the localized $a_{1g}$ defect state and the conduction-band minimum. In contrast to the $T_{1u}$ states, this excitation involves two orbitals of the same $a_{1g}$ symmetry, which leads to strong open-shell coupling and makes the optimization more challenging. For this state, we observed a collapse to a triplet-like solution in which the defect orbital remains localized at the vacancy, whereas the CBM-like orbital is delocalized over the entire supercell except at the vacancy site. This differs from the CBM obtained from a ground-state calculation, where the probability density at the vacancy site is non-zero, as shown in the SI Figure~S1. The approaches used successfully for the molecular $\pi\rightarrow\pi^\ast$ excitations in Section~\ref{sec:resA} did not yield a higher-lying singlet solution for the CBM excitation. Suppressing the open-shell coupling in the CG algorithm led to a non-stationary state, while the DIIS algorithm converged to the triplet-like state. 
Hence, a more robust approach to the eletronic structure optimization would be required to obtain the correct singlet state. For example, by explicitly enforcing the orthogonality of the excited state to the ground state~\cite{doi:10.1021/jp401323d,doi:10.1021/acs.jctc.4c01509} or via generalized mode following~\cite{Schmerwitz2023}. However, the implementation of these methods is outside the scope of the present work.
Therefore, here we report the energy of the triplet-like state. As we will discuss below, we nevertheless expect the error in the vertical excitation energy to be small, since the singlet-triplet gap is small for the excitation to the CBM.

The vertical excitation energy was calculated as
\begin{equation}
E^{\mathrm{abs}} = E^{\mathrm{ex}} - E^{\mathrm{gs}},
\end{equation}
where both energies are evaluated at the ground-state equilibrium geometry. Across all three excitations, replacing PBE by the DDH functional (PBE0 with fraction of Fock exchange $= 1/\epsilon_\infty^\text{MgO}$, see \cite{PhysRevB.89.195112}) increases the excitation energies for both ROKS and TDDFT. For ROKS, the increase ranges from 0.25 to 0.93~eV. For TDDFT, the functional dependence is larger, with increases between 0.50 and 1.81~eV. Thus, both methods show the expected upward shift upon inclusion of nonlocal exchange, but the effect is substantially stronger in TDDFT.

Comparison with embedded Bethe--Salpeter equation (BSE) calculations shows that the DDH functional generally improves the agreement with many-body perturbation theory. For the CBM excitation, BSE gives 3.7~eV using PBE reference orbitals in the same $3\times3\times3$ supercell with 215 atoms~\cite{Vorwerk2023}. ROKS-PBE underestimates this value by 0.66~eV, whereas TDDFT-PBE underestimates it by 1.36~eV. The BSE value is 4.13~eV if DDH-orbitals are used as reference, in close agreement with both ROKS-DDH and TDDFT-DDH. 

For the $^1T_{1u}$ state, the agreement with BSE also improves when DDH is used: ROKS-DDH is lower by about 0.3~eV, while TDDFT-DDH is higher by about 0.1~eV relative to the BSE result (5.23~eV with DDH-orbitals, 4.8~eV with PBE-orbitals). The $^3T_{1u}$ state remains more strongly underestimated by both methods, with deviations of about 1~eV at the DDH level (BSE results: 4.93~eV with DDH-orbitals, 4.5~eV with PBE-orbitals). 

The larger functional dependence observed for TDDFT is particularly evident for the CBM excitation, where TDDFT changes by 1.81~eV between PBE and DDH, compared with 0.93~eV for ROKS. This may be related to the Rydberg-like character of the excitation, which involves a transition from a localized vacancy state to a delocalized conduction-band state. Such excitations are known to be difficult for semilocal TDDFT and are often underestimated more severely than in $\Delta$-SCF-type approaches~\cite{A910321J,10.1063/1.3607312,doi:10.1021/jp5082802,doi:10.1021/acs.jctc.2c00160}. The weaker functional dependence of ROKS is consistent with this picture and suggests that the underlying $\Delta$-SCF description is less sensitive to the choice of the exchange-correlation functional. In addition, the collapse to the triplet-like state for the CBM excitation is less problematic for the vertical excitation energy because the singlet--triplet gap is small. For instance, using Eq.~\eqref{eq:ziegler_sum} with independently optimized unrestricted mixed and triplet configurations gives excitation energies of 3.080~eV for the singlet and 3.010~eV for the triplet state with the PBE functional, indicating that the energy error associated with the collapse is small.

We next discuss the structural relaxation in the excited states. The excited-state Franck--Condon shift is defined as
\begin{equation}
E_{\mathrm{FC,ex}} =
E^{\mathrm{ex}}_{\mathrm{eq,gs}} -
E^{\mathrm{ex}}_{\mathrm{eq,ex}},
\end{equation}
where $E^{\mathrm{ex}}_{\mathrm{eq,gs}}$ and $E^{\mathrm{ex}}_{\mathrm{eq,ex}}$ denote excited-state energies evaluated at the ground- and excited-state equilibrium geometries, respectively. The corresponding ground-state Franck--Condon shift is
\begin{equation}
E_{\mathrm{FC,gs}} =
E^{\mathrm{gs}}_{\mathrm{eq,ex}} -
E^{\mathrm{gs}}_{\mathrm{eq,gs}}.
\end{equation}
The Franck--Condon shifts are less sensitive to the functional than the vertical excitation energies, especially for ROKS. Averaged over both Franck--Condon shifts and all three excitations, the absolute PBE--DDH difference is only 0.12~eV for ROKS, compared with 0.24~eV for TDDFT. This indicates that the shape of the excited-state potential energy surface is more consistent between PBE and DDH in ROKS.

The largest discrepancy between ROKS and TDDFT occurs for the CBM excitation. At the PBE level, ROKS predicts much larger Franck--Condon shifts than TDDFT, indicating a stronger structural relaxation. Although the TDDFT Franck--Condon shifts increase substantially with DDH, a sizable difference between ROKS-DDH and TDDFT-DDH remains. This discrepancy may again be connected to the triplet-like character of the converged ROKS solution for the CBM excitation, which can lead to different excited-state forces and therefore to a different relaxed excited-state geometry.

Furthermore, we quantified structural changes between the ground and excited state equilibrium geometries by calculating mass-weighted displacements $\Delta Q$ defined as
\begin{equation}
\Delta Q =
\left[
\sum_{\alpha=1}^{N_{\mathrm{atom}}}
\sum_{i=x,y,z}
M_\alpha
\left(
R_{\alpha i}^{\mathrm{ex}} -
R_{\alpha i}^{\mathrm{gs}}
\right)^2
\right]^{1/2},
\end{equation}
where $M_\alpha$ denotes the mass of atom $\alpha$ and $R^\text{ex}_{\alpha i}$ and $R^\text{gs}_{\alpha i}$ are nucelar coordinates in excited and ground state, respectively. For ROKS, the PBE--DDH difference in $\Delta Q$ is nearly constant across the three excitations, with an average of only 0.09~amu$^{1/2}$\AA. In contrast, TDDFT shows a much stronger functional dependence, with an average PBE--DDH difference of 0.54~amu$^{1/2}$\AA. For the $T_{1u}$ states, where symmetry prevents mixing with the ground state, ROKS and TDDFT give very similar $\Delta Q$ values at the DDH level. The larger difference for the CBM excitation could be again related to the triplet-like ROKS solution for this excitation.

\begin{figure*}
    \centering
    \includegraphics[width=\textwidth]{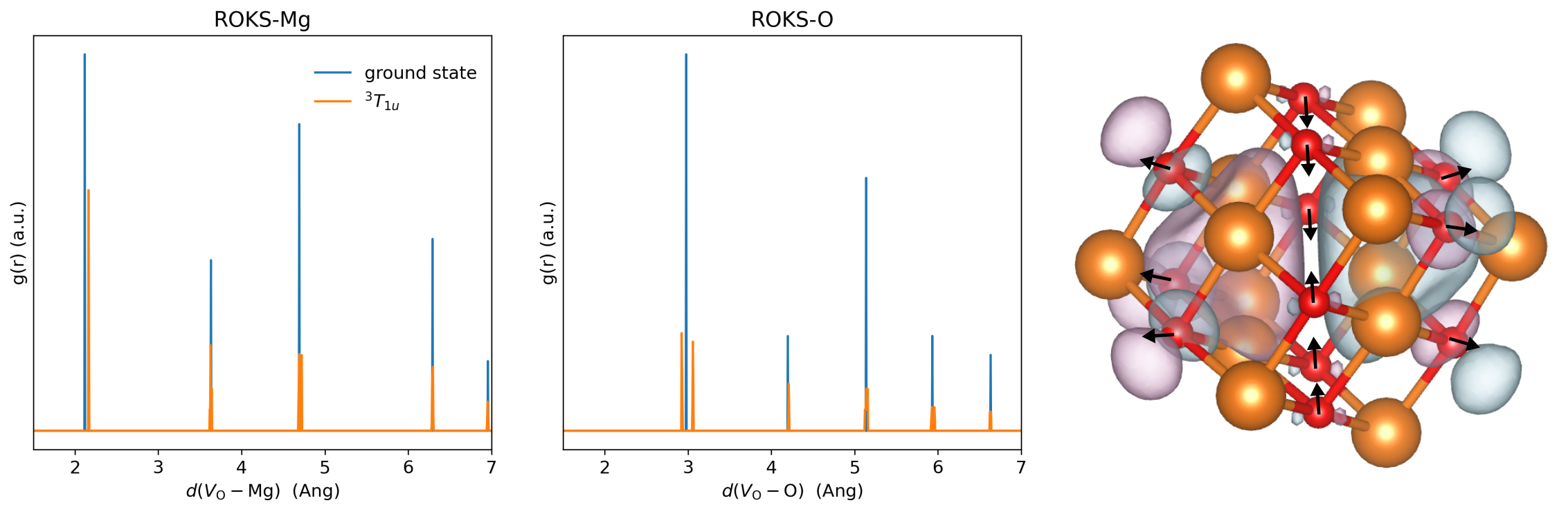}
    \caption{The radial distribution function of Mg (left) and oxygen (middle) around the oxygen vacancy in the ground state and the optimized $^3T_{1u}$-state obtained with ROKS and the DDH functional. The subfigure on the right shows the higher lying singly occupied orbital at the $^3T_{1u}$-state equilibrium structure. The arrows indicate the direction of expansion and contraction of the oxygen atoms compared to the ground state. }
    \label{fig:D3d_distortion}
\end{figure*}
We also compared radial distribution functions of Mg and O around the oxygen vacancy of the $T_{1u}$-states with the ones of the ground state to obtain more detailed insight into the differences between the structures obtained from ROKS and TDDFT with the different functionals. Figure~\ref{fig:D3d_distortion} shows the radial distribution functions for the ground state and the $^3T_{1u}$ state obtained from ROKS-DDH. The other distribution functions are reported in the SI Figures~S3 - S5.
The largest changes between ground and excited state equilibrium occur close to the vacancy site, as one would expect for a local excitation from the $a_{1g}$ to a $t_{1u}$ orbital. The Mg$^{2+}$ cations in the first coordination shell move further away from the center of the vacancy compared to the ground state. In contrast, the twelve O$^{2-}$ anions around the vacancy split into two groups in the $^3T_{1u}$-state with six anions moving closer to the vacancy and six anions moving further away. 
This causes a distortion of the structure from $O_h$ in the ground state to $D_{3d}$ symmetry in the excited state. 

The driving force behind this distortion is the minimization of the electrostatic repulsion between the electrons localized at the defect site with the bulk electrons. In the equilibrium structure of the $^3T_{1u}$-state, the higher lying singly occupied orbital is oriented along one of the diagonals of the cube formed by the atoms around the vacancy (see Figure~\ref{fig:D3d_distortion} on the right). To reduce the electrostatic interaction with the oxygen anions in the upper and lower corner of the cube along the diagonal, these oxygen atoms move further away from the center of the vacancy. The remaining six oxygen anions in the plane orthogonal to the diagonal move slightly closer to the center of the vacancy. The same split can also be observed in the TDDFT calculations (Ref.~\cite{galli_tddft}) although the $D_{3d}$ symmetry is less well respected, likely because of less careful relaxation.
For the $^1T_{1u}$-state, the same type of distortion occurs although the displacement from the ground state structure is smaller. In particular the split of the oxygen atoms into two groups is less pronounced and only clearly visible after geometry optimization of the excited state with ROKS-DDH (see SI figures S4 and S5).
In addition, the radial distribution functions reveal larger relaxation effects for TDDFT-PBE than for TDDFT-DDH or for ROKS far away from the vacancy  (see SI figures S2-S5). This could explain the large discrepancy of the $\Delta Q$-values from TDDFT-PBE with the other methods.

The emission energies were obtained from
\begin{equation}
E^{\mathrm{emi}} =
E^{\mathrm{abs}} -
E_{\mathrm{FC,ex}} -
E_{\mathrm{FC,gs}}.
\end{equation}
As for the absorption energies, DDH increases the emission energies relative to PBE for both methods. The increase is smaller for ROKS, ranging from 0.05 to 0.80~eV, than for TDDFT, where it ranges from 0.28 to 1.43~eV. For the $^1T_{1u}$ excitation, the ROKS-DDH and TDDFT-DDH emission energies agree closely because the lower ROKS vertical excitation energy is compensated by smaller Franck--Condon shifts. The $^3T_{1u}$ state shows good agreement between ROKS-DDH and TDDFT-DDH for all reported quantities. For the CBM excitation, however, the emission energies differ by more than 0.6~eV at the DDH level, despite similar vertical excitation energies, because of the different Franck--Condon shifts discussed above.

The differences between ROKS and TDDFT obtained in the present work are consistent with previous comparisons between $\Delta$-SCF and TDDFT for molecular excitations. For example, a mean absolute deviation of 0.25~eV between $\Delta$-SCF-PBE0 and TDDFT-PBE0 was reported for 16 organic dyes, with individual deviations up to 0.66~eV~\cite{Kowalczyk2011}. Another study found a mean absolute deviation of 0.547~eV between $\Delta$-SCF and TDDFT for 109 small molecules using CAM-B3LYP~\cite{BourneWorster2021}. The vertical excitation energies of MgO obtained in our study deviate between 0.02 and 0.42~eV for ROKS-DDH and TDDFT-DDH and are hence in the same range. We note that the $\Delta$-SCF calculations in  Refs~\cite{Kowalczyk2011} and \cite{BourneWorster2021} were not performed with ROKS but used Ziegler's sum rule together with unrestricted DFT calculations for the mixed and triplet states. However, ROKS and such $\Delta$-SCF approaches often yield similar results (see Figure~\ref{fig:roks_scheme}B, Ref~\cite{Frank1998}). The present results therefore suggest that the known differences between $\Delta$-SCF and TDDFT also carry over to defect excitations in solids.

Finally, the close agreement between ROKS-PBE and ROKS-DDH for the Franck--Condon shifts and $\Delta Q$ values is encouraging for efficient excited-state simulations. It indicates a large phase-space overlap between the PBE and DDH excited-state potential energy surfaces. This is favorable for $\Delta$-learning machine-learning strategies~\cite{doi:10.1021/acs.jctc.5b00099}, where a large set of excited-state structures can be generated at the inexpensive PBE level and a smaller subset can be recomputed with the more accurate hybrid functional to train a correction model.

\section{Conclusion}
Variational excited-state DFT enables the modeling of electronically excited states at a computational cost comparable to ground-state calculations, making it attractive for large systems and long time scales. However, single-configuration approaches are generally spin-contaminated and may yield inaccurate excitation energies.
Spin-pure excitation energies can be recovered by variationally minimizing a weighted combination of mixed-spin and triplet configurations, as realized in restricted open-shell DFT (ROKS).

In this work, we implemented ROKS within the plane-wave projector augmented-wave framework of VASP using a preconditioned conjugate-gradient algorithm and a DIIS solver and derived the corresponding analytical forces. Benchmark calculations for eight organic molecules show excellent agreement with Q-Chem, with mean deviations of about $30\,\mathrm{meV}$, consistent with expected basis-set differences.

We further benchmarked our implementation by modeling the three lowest lying excited states of MgO with a neutral oxygen vacancy and comparing them with results from TDDFT~\cite{galli_tddft}. 
The vertical excitation energies differ between 0.02-0.42 between ROKS and TDDFT if evaluated with a dielectric-dependent hybrid DFT functional similar to deviations between $\Delta$-SCF and TDDFT reported for molecular systems~\cite{Kowalczyk2011,BourneWorster2021}. 
The Franck-Condon shifts deviate on average by 0.14~eV between the two methods and mass-weighted displacements by 0.12~amu$^{1/2}$\AA. In addition, the ROKS results are less sensitive to the choice of the DFT-functional. Specifically, if these properties are evaluated using the PBE functional, the agreement with results obtained from the hybrid functional is significantly better for ROKS than for TDDFT.

Overall, our results demonstrate that ROKS provides excitation energies of similar quality to TDDFT while retaining the favorable scaling of ground-state DFT. This makes the approach particularly promising for excited-state calculations in extended systems.


\section*{Supplementary Information}
The supplementary information provides derivations of the expression for the Lagrange multipliers and of the contribution of the PAW overlap term to the forces. It contains also figures of the shape of the defect orbital and the CBM in MgO with an oxygen vacancy and
figures of the radial distribution functions around the oxygen vacancy in MgO for the $^1T_{1u}$ and $^3T_{1u}$ states.

\section*{Data Availability}
Input and output files of the VASP calculations for MgO with an oxygen vacancy and the molecules discussed in Section~\ref{sec:resA} are available on Zenodo at \doi{https://doi.org/10.5281/zenodo.20326636}.

\begin{acknowledgments}
The authors acknowledge funding from the Austrian Science Fund (FWF) for the Cluster of Excellence “Materials for Energy Conversion and Storage” [grant 10.55776/COE5, accessible via \url{//www.fwf.ac.at/en/discover/research-radar}] as well as the Austrian Scientific Computing (ASC) for generous allocation of computational time and the University of Vienna for
continuous support.
\end{acknowledgments}

\section*{AUTHOR DECLARATIONS}
\subsection*{Conflict of Interest}
The authors have no conflicts to disclose.

\FloatBarrier
\bibliography{ref}

@article{Ziegler1977,
  author = {Tom Ziegler and Arvi Rank and Evert J Baerends},
  title = {{On the Calculation of Multiplet Energies by the Hartree-Fock-Slater Method}},
  journal = {Theor. Chim. Acta},
  volume = {43},
  pages = {261--271},
  year = {1977}
}

@article{Filatov1998,
  author = {Michael Filatov and Sason Shaik},
  title = {{Spin-Restricted Density Functional Approach to the Open-Shell Problem}},
  journal = {Chem. Phys. Lett.},
  volume = {288},
  pages = {689--697},
  year = {1998}
}

@article{Frank1998,
  author = {Irmgard Frank and Jürg Hutter and Dominik Marx and Michele Parrinello},
  title = {{Molecular Dynamics in Low-Spin Excited States}},
  journal = {J. Chem. Phys.},
  volume = {108},
  number = {10},
  pages = {4060--4069},
  year = {1998},
  month = {3},
  doi = {10.1063/1.475804}
}

@article{Grimm2003,
  author = {Stephan Grimm and Christel Nonnenberg and Irmgard Frank},
  title = {{Restricted Open-Shell Kohn-Sham Theory for $\phi - \pi^*$ Transitions. I. Polyenes, Cyanines, and Protonated Imines}},
  journal = {J. Chem. Phys.},
  volume = {119},
  number = {22},
  pages = {11574--11584},
  year = {2003},
  month = {12},
  doi = {10.1063/1.1623742}
}

@article{FRIEDRICHS200817,
  author = {Jana Friedrichs and Konstantina Damianos and Irmgard Frank},
  title = {{Solving Restricted Open-Shell Equations in Excited State Molecular Dynamics Simulations}},
  journal = {Chem. Phys.},
  volume = {347},
  number = {1},
  pages = {17--24},
  year = {2008},
  doi = {10.1016/j.chemphys.2007.09.035},
  note = {Ultrafast Photoinduced Processes in Polyatomic Molecules}
}

@article{Hirao1973,
  author = {K. Hirao and H. Nakatsuji},
  title = {{General SCF Operator Satisfying Correct Variational Condition}},
  journal = {J. Chem. Phys.},
  volume = {59},
  number = {3},
  pages = {1457--1462},
  year = {1973},
  doi = {10.1063/1.1680203}
}

@article{Kresse1996,
  author = {Georg Kresse and J. Furthmüller},
  title = {{Efficiency of Ab-Initio Total Energy Calculations for Metals and Semiconductors using a Plane-Wave Basis Set}},
  journal = {Comput. Mater. Sci.},
  volume = {6},
  pages = {15--50},
  year = {1996}
}

@article{PhysRevB.54.11169,
  author = {Kresse, G. and Furthm\"uller, J.},
  title = {{Efficient Iterative Schemes for Ab Initio Total-Energy Calculations using a Plane-Wave Basis Set}},
  journal = {Phys. Rev. B},
  volume = {54},
  number = {16},
  pages = {11169--11186},
  year = {1996},
  month = {Oct},
  doi = {10.1103/PhysRevB.54.11169}
}

@article{PhysRevB.59.1758,
  author = {Kresse, G. and Joubert, D.},
  title = {{From Ultrasoft Pseudopotentials to the Projector Augmented-Wave Method}},
  journal = {Phys. Rev. B},
  volume = {59},
  number = {3},
  pages = {1758--1775},
  year = {1999},
  month = {Jan},
  doi = {10.1103/PhysRevB.59.1758}
}

@article{Kowalczyk2013,
  author = {Tim Kowalczyk and Takashi Tsuchimochi and Po Ta Chen and Laken Top and Troy Van Voorhis},
  title = {{Excitation Energies and Stokes Shifts from a Restricted Open-Shell Kohn-Sham Approach}},
  journal = {J. Chem. Phys.},
  volume = {138},
  number = {16},
  year = {2013},
  month = {4},
  doi = {10.1063/1.4801790}
}

@article{DanielEdwards1987,
  author = {W Daniel Edwards and Michael C Zerner},
  title = {{A Generalized Restricted Open-Shell Fock Operator*}},
  journal = {Theor. Chim. Acta},
  volume = {72},
  pages = {347--361},
  year = {1987}
}

@article{Goedecker1992,
  author = {S Goedecker and K Maschke},
  title = {{Operator Approach in the Linearized Augmented-Plane-Wave Method: Efficient Electronic-Structure Calculations Including Forces}},
  journal = {Phys. Rev. B},
  volume = {45},
  pages = {15--1992},
  year = {1992}
}

@article{10.1063/1.1926272,
  author = {Paier, Joachim and Hirschl, Robin and Marsman, Martijn and Kresse, Georg},
  title = {{The Perdew–Burke–Ernzerhof Exchange-Correlation Functional Applied to the G2-1 Test Set using a Plane-Wave Basis Set}},
  journal = {J. Chem. Phys.},
  volume = {122},
  number = {23},
  pages = {234102},
  year = {2005},
  month = {06},
  doi = {10.1063/1.1926272}
}

@article{galli_tddft,
  author = {Jin, Yu and Yu, Victor Wen-zhe and Govoni, Marco and Xu, Andrew C. and Galli, Giulia},
  title = {{Excited State Properties of Point Defects in Semiconductors and Insulators Investigated with Time-Dependent Density Functional Theory}},
  journal = {J. Chem. Theory Comput.},
  volume = {19},
  number = {23},
  pages = {8689--8705},
  year = {2023},
  doi = {10.1021/acs.jctc.3c00986}
}

@article{10.1063/1.3607312,
  author = {Yang, Ke and Peverati, Roberto and Truhlar, Donald G. and Valero, Rosendo},
  title = {{Density Functional Study of Multiplicity-Changing Valence and Rydberg Excitations of P-Block Elements: Delta Self-Consistent Field, Collinear Spin-Flip Time-Dependent Density Functional Theory (DFT), and Conventional Time-Dependent DFT}},
  journal = {J. Chem. Phys.},
  volume = {135},
  number = {4},
  pages = {044118},
  year = {2011},
  month = {07},
  doi = {10.1063/1.3607312}
}

@article{doi:10.1021/jp5082802,
  author = {Seidu, Issaka and Krykunov, Mykhaylo and Ziegler, Tom},
  title = {{Applications of Time-Dependent and Time-Independent Density Functional Theory to Rydberg Transitions}},
  journal = {J. Phys. Chem. A},
  volume = {119},
  number = {21},
  pages = {5107--5116},
  year = {2015},
  doi = {10.1021/jp5082802}
}

@article{A910321J,
  author = {Tozer, David J. and Handy, Nicholas C.},
  title = {{On the Determination of Excitation Energies using Density Functional Theory}},
  journal = {Phys. Chem. Chem. Phys.},
  volume = {2},
  number = {10},
  pages = {2117--2121},
  year = {2000},
  doi = {10.1039/A910321J}
}

@article{doi:10.1021/acs.jctc.2c00160,
  author = {Liang, Jiashu and Feng, Xintian and Hait, Diptarka and Head-Gordon, Martin},
  title = {{Revisiting the Performance of Time-Dependent Density Functional Theory for Electronic Excitations: Assessment of 43 Popular and Recently Developed Functionals from Rungs One to Four}},
  journal = {J. Chem. Theory Comput.},
  volume = {18},
  number = {6},
  pages = {3460--3473},
  year = {2022},
  doi = {10.1021/acs.jctc.2c00160}
}

@article{Vorwerk2023,
  author = {Christian Vorwerk and Giulia Galli},
  title = {{Disentangling Photoexcitation and Photoluminescence Processes in Defective MgO}},
  journal = {Phys. Rev. Mater.},
  volume = {7},
  number = {3},
  year = {2023},
  month = {3},
  doi = {10.1103/PhysRevMaterials.7.033801}
}

@article{Schmerwitz2023,
  author = {Yorick L.A. Schmerwitz and Gianluca Levi and Hannes Jónsson},
  title = {{Calculations of Excited Electronic States by Converging on Saddle Points using Generalized Mode Following}},
  journal = {J. Chem. Theory Comput.},
  volume = {19},
  number = {12},
  pages = {3634--3651},
  year = {2023},
  month = {6},
  doi = {10.1021/acs.jctc.3c00178}
}

@article{Levi2020,
  author = {Gianluca Levi and Aleksei V. Ivanov and Hannes Jónsson},
  title = {{Variational Calculations of Excited States: Via Direct Optimization of the Orbitals in DFT}},
  journal = {Faraday Discuss.},
  volume = {224},
  pages = {448--466},
  year = {2020},
  month = {12},
  doi = {10.1039/d0fd00064g}
}

@article{doi:10.1021/acs.jctc.9b01127,
  author = {Hait, Diptarka and Head-Gordon, Martin},
  title = {{Excited State Orbital Optimization via Minimizing the Square of the Gradient: General Approach and Application to Singly and Doubly Excited States via Density Functional Theory}},
  journal = {J. Chem. Theory Comput.},
  volume = {16},
  number = {3},
  pages = {1699--1710},
  year = {2020},
  doi = {10.1021/acs.jctc.9b01127}
}

@article{Kowalczyk2011,
  author = {Tim Kowalczyk and Shane R. Yost and Troy Van Voorhis},
  title = {{Assessment of the $\Delta$-SCF Density Functional Theory Approach for Electronic Excitations in Organic Dyes}},
  journal = {J. Chem. Phys.},
  volume = {134},
  number = {5},
  year = {2011},
  month = {2},
  doi = {10.1063/1.3530801}
}

@article{BourneWorster2021,
  author = {Susannah Bourne Worster and Oliver Feighan and Frederick R. Manby},
  title = {{Reliable Transition Properties from Excited-State Mean-Field Calculations}},
  journal = {J. Chem. Phys.},
  volume = {154},
  number = {12},
  year = {2021},
  month = {3},
  doi = {10.1063/5.0041233}
}

@incollection{SLATER19721,
  author = {John C. Slater},
  title = {{Statistical Exchange-Correlation in the Self-Consistent Field}},
  series = {Advances in Quantum Chemistry},
  editor = {Per-Olov Löwdin},
  publisher = {Academic Press},
  volume = {6},
  pages = {1--92},
  year = {1972},
  doi = {10.1016/S0065-3276(08)60541-9},
  url = {https://www.sciencedirect.com/science/article/pii/S0065327608605419},
  issn = {0065-3276}
}

@article{PhysRevB.78.075441,
  author = {Gavnholt, Jeppe and Olsen, Thomas and Engelund, Mads and Schi\o{}tz, Jakob},
  title = {{$\ensuremath{\Delta}$ Self-Consistent Field Method to Obtain Potential Energy Surfaces of Excited Molecules on Surfaces}},
  journal = {Phys. Rev. B},
  volume = {78},
  number = {7},
  pages = {075441},
  year = {2008},
  month = {Aug},
  doi = {10.1103/PhysRevB.78.075441}
}

@article{Hait2021,
  author = {Diptarka Hait and Martin Head-Gordon},
  title = {{Orbital Optimized Density Functional Theory for Electronic Excited States}},
  journal = {J. Phys. Chem. Lett.},
  volume = {12},
  number = {19},
  pages = {4517--4529},
  year = {2021},
  month = {5},
  doi = {10.1021/acs.jpclett.1c00744}
}

@article{PhysRevB.77.155206,
  author = {Gali, Adam and Fyta, Maria and Kaxiras, Efthimios},
  title = {{Ab Initio Supercell Calculations on Nitrogen-Vacancy Center in Diamond: Electronic Structure and Hyperfine Tensors}},
  journal = {Phys. Rev. B},
  volume = {77},
  number = {15},
  pages = {155206},
  year = {2008},
  month = {Apr},
  doi = {10.1103/PhysRevB.77.155206}
}

@article{Alkauskas_2014,
  author = {Alkauskas, Audrius and Buckley, Bob B and Awschalom, David D and Van de Walle, Chris G},
  title = {{First-Principles Theory of the Luminescence Lineshape for the Triplet Transition in Diamond NV Centres}},
  journal = {New J. Phys.},
  volume = {16},
  number = {7},
  pages = {073026},
  year = {2014},
  month = {jul},
  doi = {10.1088/1367-2630/16/7/073026}
}

@article{Ivanov2023,
  author = {Aleksei V. Ivanov and Yorick L.A. Schmerwitch and Gianluca Levi and Hannes Jónsson},
  title = {{Electronic Excitations of the Charged Nitrogen-Vacancy Center in Diamond Obtained using Time-Independent Variational Density Functional Calculations}},
  journal = {SciPost Phys.},
  volume = {15},
  number = {1},
  year = {2023},
  month = {7},
  doi = {10.21468/SciPostPhys.15.1.009}
}

@article{PhysRevB.96.085204,
  author = {Cs\'or\'e, A. and von Bardeleben, H. J. and Cantin, J. L. and Gali, A.},
  title = {{Characterization and Formation of NV Centers in $3C, 4H$, And $6H$ SiC: An Ab Initio Study}},
  journal = {Phys. Rev. B},
  volume = {96},
  number = {8},
  pages = {085204},
  year = {2017},
  month = {Aug},
  doi = {10.1103/PhysRevB.96.085204}
}

@article{PhysRevMaterials.7.096202,
  author = {Mohseni, Meysam and Udvarhelyi, P\'eter and Thiering, Gerg\ifmmode \mbox{\H{o}}\else \H{o}\fi{} and Gali, Adam},
  title = {{Positively Charged Carbon Vacancy Defect as a Near-Infrared Emitter in 4H-SiC}},
  journal = {Phys. Rev. Mater.},
  volume = {7},
  number = {9},
  pages = {096202},
  year = {2023},
  month = {Sep},
  doi = {10.1103/PhysRevMaterials.7.096202}
}

@article{https://doi.org/10.1002/adma.202408424,
  author = {Aberl, Johannes and Navarrete, Enrique Prado and Karaman, Merve and Enriquez, Diego Haya and Wilflingseder, Christoph and Salomon, Andreas and Primetzhofer, Daniel and Schubert, Markus Andreas and Capellini, Giovanni and Fromherz, Thomas and Deák, Peter and Udvarhelyi, Peter and Li, Song and Gali, Adam and Brehm, Moritz},
  title = {{All-Epitaxial Self-Assembly of Silicon Color Centers Confined within Sub-Nanometer Thin Layers using Ultra-Low Temperature Epitaxy}},
  journal = {Adv. Mater.},
  volume = {36},
  number = {48},
  pages = {2408424},
  year = {2024},
  doi = {10.1002/adma.202408424}
}

@article{Cholsuk2024,
  author = {Chanaprom Cholsuk and Ashkan Zand and Aslı Çakan and Tobias Vogl},
  title = {{The hBN Defects Database: A Theoretical Compilation of Color Centers in Hexagonal Boron Nitride}},
  journal = {J. Phys. Chem. C},
  volume = {128},
  number = {30},
  pages = {12716--12725},
  year = {2024},
  month = {8},
  doi = {10.1021/acs.jpcc.4c03404}
}

@article{PhysRevLett.123.127401,
  author = {Turiansky, Mark E. and Alkauskas, Audrius and Bassett, Lee C. and Van de Walle, Chris G.},
  title = {{Dangling Bonds in Hexagonal Boron Nitride as Single-Photon Emitters}},
  journal = {Phys. Rev. Lett.},
  volume = {123},
  number = {12},
  pages = {127401},
  year = {2019},
  month = {Sep},
  doi = {10.1103/PhysRevLett.123.127401}
}

@article{https://doi.org/10.1002/adom.202500593,
  author = {Gale, Angus and Kianinia, Mehran and Horder, Jake and Tweedie, Connor and Singhal, Mridul and Scognamiglio, Dominic and Qi, Jiajie and Liu, Kaihui and Verdi, Carla and Aharonovich, Igor and Toth, Milos},
  title = {{Quantum Emitters in Rhombohedral Boron Nitride}},
  journal = {Adv. Opt. Mater.},
  volume = {13},
  number = {27},
  pages = {e00593},
  year = {2025},
  doi = {10.1002/adom.202500593}
}

@article{doi:10.1021/jp801738f,
  author = {Gilbert, Andrew T. B. and Besley, Nicholas A. and Gill, Peter M. W.},
  title = {{Self-Consistent Field Calculations of Excited States using the Maximum Overlap Method (MOM)}},
  journal = {J. Phys. Chem. A},
  volume = {112},
  number = {50},
  pages = {13164--13171},
  year = {2008},
  doi = {10.1021/jp801738f}
}

@article{doi:10.1021/acs.jctc.7b00994,
  author = {Barca, Giuseppe
M. J. and Gilbert, Andrew T. B. and Gill, Peter M. W.},
  title = {{Simple Models for Difficult Electronic Excitations}},
  journal = {J. Chem. Theory Comput.},
  volume = {14},
  number = {3},
  pages = {1501--1509},
  year = {2018},
  doi = {10.1021/acs.jctc.7b00994}
}

@article{doi:10.1021/acs.jctc.0c00502,
  author = {Carter-Fenk, Kevin and Herbert, John M.},
  title = {{State-Targeted Energy Projection: A Simple and Robust Approach to Orbital Relaxation of Non-Aufbau Self-Consistent Field Solutions}},
  journal = {J. Chem. Theory Comput.},
  volume = {16},
  number = {8},
  pages = {5067--5082},
  year = {2020},
  doi = {10.1021/acs.jctc.0c00502}
}

@article{Hait2020,
  author = {Diptarka Hait and Martin Head-Gordon},
  title = {{Excited State Orbital Optimization via Minimizing the Square of the Gradient: General Approach and Application to Singly and Doubly Excited States via Density Functional Theory}},
  journal = {J. Chem. Theory Comput.},
  volume = {16},
  number = {3},
  pages = {1699--1710},
  year = {2020},
  month = {3},
  doi = {10.1021/acs.jctc.9b01127}
}

@article{Filatov1999,
  author = {Michael Filatov and Sason Shaik},
  title = {{Application of Spin-Restricted Open-Shell Kohn-Sham Method to Atomic and Molecular Multiplet States}},
  journal = {J. Chem. Phys.},
  volume = {110},
  number = {1},
  pages = {116--125},
  year = {1999},
  doi = {10.1063/1.477941}
}

@article{C9CP06419B,
  author = {Schwermann, Christian and Doltsinis, Nikos L.},
  title = {{Exciton Transfer Free Energy from Car–Parrinello Molecular Dynamics}},
  journal = {Phys. Chem. Chem. Phys.},
  volume = {22},
  number = {19},
  pages = {10526--10535},
  year = {2020},
  doi = {10.1039/C9CP06419B}
}

@article{D3CP00533J,
  author = {Diarra, Cheick Oumar and Boero, Mauro and Steveler, Emilie and Heiser, Thomas and Martin, Evelyne},
  title = {{Exciton Diffusion in Poly(3-Hexylthiophene) by First-Principles Molecular Dynamics}},
  journal = {Phys. Chem. Chem. Phys.},
  volume = {25},
  number = {22},
  pages = {15539--15546},
  year = {2023},
  doi = {10.1039/D3CP00533J}
}

@article{PhysRevB.101.100101,
  author = {Fedorov, Ilya D. and Orekhov, Nikita D. and Stegailov, Vladimir V.},
  title = {{Nonadiabatic Effects and Excitonlike States During the Insulator-to-Metal Transition in Warm Dense Hydrogen}},
  journal = {Phys. Rev. B},
  volume = {101},
  number = {10},
  pages = {100101},
  year = {2020},
  month = {Mar},
  doi = {10.1103/PhysRevB.101.100101}
}

@article{10.1063/5.0288340,
  author = {Fominykh, Nikita A. and Stegailov, Vladimir V.},
  title = {{Exciton Diffusion in MoS2 Monolayer from First-Principles Molecular Dynamics}},
  journal = {J. Chem. Phys.},
  volume = {163},
  number = {11},
  pages = {114704},
  year = {2025},
  month = {09},
  doi = {10.1063/5.0288340}
}

@article{Roothaan1960,
  author = {C. C. J. Roothaan},
  title = {{Self-Consistent Field Theory for Open Shells of Electronic Systems}},
  journal = {Rev. Mod. Phys.},
  volume = {32},
  number = {2},
  pages = {179--185},
  year = {1960},
  month = {4},
  doi = {10.1103/RevModPhys.32.179}
}

@article{Jacobi,
  author = {C.G.J. Jacobi},
  title = {{Über ein leichtes Verfahren die in der Theorie der Säcularstörungen vorkommenden Gleichungen numerisch aufzulösen*).}},
  journal = {J. Reine Angew. Math.},
  volume = {1846},
  number = {30},
  pages = {51--94},
  year = {1846},
  doi = {10.1515/crll.1846.30.51}
}

@article{PhysRevB.50.17953,
  author = {Bl\"ochl, P. E.},
  title = {{Projector Augmented-Wave Method}},
  journal = {Phys. Rev. B},
  volume = {50},
  number = {24},
  pages = {17953--17979},
  year = {1994},
  month = {Dec},
  doi = {10.1103/PhysRevB.50.17953}
}

@article{shao2015,
  author = {Shao, Yihan and Gan, Zhengting and Epifanovsky, Evgeny and Gilbert, Andrew TB and Wormit, Michael and Kussmann, Joerg and Lange, Adrian W and Behn, Andrew and Deng, Jia and Feng, Xintian and others},
  title = {{Advances in Molecular Quantum Chemistry Contained in the Q-Chem 4 Program Package}},
  journal = {Mol. Phys.},
  volume = {113},
  number = {2},
  pages = {184--215},
  year = {2015}
}

@article{epifanovsky2021,
  author = {Epifanovsky, Evgeny and Gilbert, Andrew TB and Feng, Xintian and Lee, Joonho and Mao, Yuezhi and Mardirossian, Narbe and Pokhilko, Pavel and White, Alec F and Coons, Marc P and Dempwolff, Adrian L and others},
  title = {{Software for the Frontiers of Quantum Chemistry: An Overview of Developments in the Q-Chem 5 Package}},
  journal = {J. Chem. Phys.},
  volume = {155},
  number = {8},
  year = {2021}
}

@article{Hutter1994,
  author = {Jürg Hutter and Hans Peter Lüthi and Michele Parrinello},
  title = {{Electronic Structure Optimization in Plane-Wave-Based Density Functional Calculations by Direct Inversion in the Iterative Subspace}},
  journal = {Comput. Mater. Sci.},
  volume = {2},
  number = {2},
  pages = {244--248},
  year = {1994},
  month = {3},
  doi = {10.1016/0927-0256(94)90105-8}
}

@book{Carbo1978,
  author = {Carb{\'{o}}, Ramon and Riera, Joseph M.},
  title = {{A General SCF Theory}},
  series = {Lecture Notes in Chemistry},
  publisher = {Springer Berlin, Heidelberg},
  volume = {5},
  year = {1978},
  doi = {10.1007/978-3-642-93075-1},
  isbn = {978-3-642-93075-1}
}

@article{doi:10.1021/acs.jctc.4c01509,
  author = {Pham, Hanh D. M. and Khaliullin, Rustam Z.},
  title = {{Direct Unconstrained Optimization of Excited States in Density Functional Theory}},
  journal = {J. Chem. Theory Comput.},
  volume = {21},
  number = {8},
  pages = {3902--3912},
  year = {2025},
  doi = {10.1021/acs.jctc.4c01509}
}

@article{doi:10.1021/jp401323d,
  author = {Evangelista, Francesco A. and Shushkov, Philip and Tully, John C.},
  title = {{Orthogonality Constrained Density Functional Theory for Electronic Excited States}},
  journal = {J. Phys. Chem. A},
  volume = {117},
  number = {32},
  pages = {7378--7392},
  year = {2013},
  doi = {10.1021/jp401323d}
}

@article{doi:10.1021/acs.jctc.5b00099,
  author = {Ramakrishnan, Raghunathan and Dral, Pavlo O. and Rupp, Matthias and von Lilienfeld, O. Anatole},
  title = {{Big Data Meets Quantum Chemistry Approximations: The $\Delta$-Machine Learning Approach}},
  journal = {J. Chem. Theory Comput.},
  volume = {11},
  number = {5},
  pages = {2087--2096},
  year = {2015},
  doi = {10.1021/acs.jctc.5b00099}
}

@book{Robin1985HigherExcitedStatesVol3,
  author = {Robin, M. B.},
  title = {{Higher Excited States of Polyatomic Molecules}},
  publisher = {Academic},
  address = {New York},
  volume = {3},
  year = {1985}
}

@article{doi:10.1021/acs.jctc.3c00089,
  author = {Rossomme, Elliot and Cunha, Leonardo A. and Li, Wanlu and Chen, Kaixuan and McIsaac, Alexandra R. and Head-Gordon, Teresa and Head-Gordon, Martin},
  title = {{The Good, the Bad, and the Ugly: Pseudopotential Inconsistency Errors in Molecular Applications of Density Functional Theory}},
  journal = {J. Chem. Theory Comput.},
  volume = {19},
  number = {10},
  pages = {2827--2841},
  year = {2023},
  doi = {10.1021/acs.jctc.3c00089}
}

@article{PhysRevB.89.195112,
  author = {Skone, Jonathan H. and Govoni, Marco and Galli, Giulia},
  title = {{Self-Consistent Hybrid Functional for Condensed Systems}},
  journal = {Phys. Rev. B},
  volume = {89},
  number = {19},
  pages = {195112},
  year = {2014},
  month = {May},
  doi = {10.1103/PhysRevB.89.195112}
}

@article{10.1093/comjnl/7.2.149,
  author = {Fletcher, R. and Reeves, C. M.},
  title = {{Function Minimization by Conjugate Gradients}},
  journal = {Comput. J.},
  volume = {7},
  number = {2},
  pages = {149--154},
  year = {1964},
  month = {01},
  doi = {10.1093/comjnl/7.2.149}
}

@article{PhysRevB.40.12255,
  author = {Teter, Michael P. and Payne, Michael C. and Allan, Douglas C.},
  title = {{Solution of Schr\"odinger's Equation for Large Systems}},
  journal = {Phys. Rev. B},
  volume = {40},
  number = {18},
  pages = {12255--12263},
  year = {1989},
  month = {Dec},
  doi = {10.1103/PhysRevB.40.12255}
}

@article{molecules29184509,
  author = {Büchel, Ralf and Álvarez, Luis and Grage, Jan and Maniscalco, Dominykas and Frank, Irmgard},
  title = {{On the Simulation of Photoreactions using Restricted Open-Shell Kohn–Sham Theory}},
  journal = {Molecules},
  volume = {29},
  number = {18},
  article-number = {4509},
  year = {2024},
  doi = {10.3390/molecules29184509},
  articleno = {4509},
  pubmedid = {39339507}
}

\end{document}


\renewcommand{\thesection}{S\arabic{section}}
\renewcommand{\thefigure}{S\arabic{figure}}
\renewcommand{\thetable}{S\arabic{table}}

\setcounter{section}{0}
\setcounter{figure}{0}
\setcounter{table}{0}

\title{Supplementary Information: \\ 
Excited States from Restricted Open Shell Plane-Wave DFT}

\author{Michael J. Sahre*}
\affiliation{University of Vienna, Faculty of Physics, Kolingasse 14, A-1090 Vienna, Austria}

\author{Marco Romanelli}
\affiliation{Institute of Theoretical Chemistry, Faculty of Chemistry, University of Vienna, W\"ahringer Str. 17, 1090 Vienna, Austria}

\author{Martijn Marsman}
\affiliation{VASP Software GmbH, Berggasse 21, A-1090 Vienna, Austria}

\author{Leticia Gonz\'{a}lez}
\affiliation{Institute of Theoretical Chemistry, Faculty of Chemistry, University of Vienna, W\"ahringer Str. 17, 1090 Vienna, Austria}

\author{Georg Kresse}
\affiliation{University of Vienna, Faculty of Physics, Kolingasse 14, A-1090 Vienna, Austria}
\affiliation{VASP Software GmbH, Berggasse 21, A-1090 Vienna, Austria}

\maketitle

\tableofcontents

\section{Derivation of the Lagrange multipliers}
Starting from the energy minimum condition
\begin{equation}\label{eq:stationary_cond}
    dE = \sum_n \braket{dn}{g_n} + \braket{g_n}{dn} = 0,
\end{equation}
the orbital variation $\ket{d n}$ can be expressed in a complete basis $\ket{m}$ as
\begin{equation}\label{eq:def_dphi}
    \ket{d n} = \sum_m \eta_{mn} \ket{m} = \sum_m (h_{mn} + a_{mn}) \ket{m},
\end{equation}
where the matrix element $\eta_{mn}$ is decomposed into the hermitian matrix $h_{mn} = \frac{1}{2} (\eta_{mn}+\eta_{nm}^*) $ and the anti-hermitian matrix $a_{mn} = \frac{1}{2}(\eta_{mn} - \eta_{nm}^*)$.

Insertion of this definition in Eq.~\eqref{eq:stationary_cond} and switching of the summation order yields
\begin{equation}
\begin{split}
    dE &= \sum_{m,n} \eta_{mn}^* \braket{m}{g_n} + \eta_{nm} \braket{g_m}{n} \\
    &= \sum_{m,n} h_{mn}^* \braket{m}{g_n} + h_{nm} \braket{g_m}{n} + a_{mn}^* \braket{m}{g_n} + a_{nm} \braket{g_m}{n}
\end{split}
\end{equation}

Since $h_{mn}^* = h_{nm}$ and $a_{mn}^* = -a_{nm}$
\begin{equation}
        dE = \sum_{n,m} h_{nm} (\braket{g_m}{n} + \braket{m}{g_n}) + a_{nm} (\braket{g_m}{n} - \braket{m}{g_n}) ,
\end{equation}
such that the energy variation $dE = 0$ if,
\begin{subequations}\label{eq:gamma_condition}
\begin{align}
    \braket{g_m}{n} + \braket{m}{g_n} &= 0 \label{eq:gamma_condition_a} \\
    \braket{g_m}{n} - \braket{m}{g_n} &= 0 \label{eq:gamma_condition_b}
\end{align}
\end{subequations}

Inserting the definition of the gradient $\ket{g_n}$ into Eq.~\eqref{eq:gamma_condition_a} and Eq.~\eqref{eq:gamma_condition_b}, respectively, leads to
\begin{subequations}\label{eq:gamma_condition2}
\begin{align}
    \sum_{L,\sigma} (f_{n,L}^\sigma + f_{m,L}^\sigma) c_L \mel{m}{H_L^\sigma}{n} = \gamma_{nm} + \gamma_{mn}^* \label{eq:gamma_condition_2a} \\
    \sum_{L,\sigma} (f_{m,L}^\sigma - f_{n,L}^\sigma) c_L \mel{m}{H_L^\sigma}{n} = \gamma_{mn}^* -  \gamma_{nm} \label{eq:gamma_condition_2b}
\end{align}
\end{subequations}

By choosing $\gamma$ to be hermitian, we finally obtain
\begin{equation}\label{eq:gamma_def}
    \gamma_{nm} = \frac{1}{2} \sum_{L,\sigma} (f_{n,L}^\sigma + f_{m,L}^\sigma) c_L \mel{m}{H_L^\sigma}{n}
\end{equation}
from Eq.~\eqref{eq:gamma_condition_2a}, while Eq.~\eqref{eq:gamma_condition_2b} is fulfilled for any hermitian $\gamma$ since $\gamma_{mn}^* -  \gamma_{nm} = 0 $ and hence  Eq.~\eqref{eq:gamma_condition_2b} simplifies to
\begin{equation}\label{eq:gamma_def_antih}
    \sum_{L,\sigma} (f_{m,L}^\sigma - f_{n,L}^\sigma) c_L \mel{m}{H_L^\sigma}{n} = 0.
\end{equation}
The second condition defined by Eqs.~\eqref{eq:gamma_condition_b} or \eqref{eq:gamma_condition_2b} is independent of $\gamma$ since the corresponding orbital variations are generated by an anti-hermitian matrix, i.e. they correspond to first order to a unitary transformation. Thus, the resulting orbitals $\ket{n + dn}$, $\ket{m + dm}$ are to first order always orthogonal to each other and no Lagrange multiplier "is needed" to maintain the constraint. Furthermore, the equation is automatically fulfilled if $f_n = f_m$.
However, for open shell systems $f_n \neq f_m \neq 0$ if $n$ and $m$ belong to different shells. Hence, in such a case one must ensure that both equations, \eqref{eq:gamma_def} and \eqref{eq:gamma_def_antih}, are fulfilled.\cite{Hirao1973}


\section{Transformation of the overlap term}
The overlap term in the force expression Eq.~(24) can be expressed as
\begin{equation}
    \sum_{n,m} \bar{\gamma}_{nm} S'_{mn} = \Tr{\bar{\gamma} \cdot S'}
\end{equation}
with $S'_{mn} = \mel{m}{\frac{\partial S}{\partial R}}{n}$ and $\bar{\gamma}$ being the complex conjugate of $\gamma$.
We can further rewrite this trace as
\begin{equation}
    \Tr{\bar{\gamma} S'} = \Tr{\bar{\gamma} U U^\dagger S' U U^\dagger} = \Tr{ U^\dagger \bar{\gamma} U U^\dagger S' U}
\end{equation}
by introducing a unitary matrix $U$ and exploiting that the trace of a matrix product is invariant under circular shifts.
By choosing the matrix $U$ such that it diagonalizes $\bar{\gamma}$ and defining
\begin{equation}
    \Lambda = U^\dagger \bar{\gamma} U
\end{equation}
and 
\begin{equation}
    \tilde{S}' = U^\dagger S' U
\end{equation}
we obtain
\begin{equation}
    \Tr{\bar{\gamma} S'} = \Tr{\Lambda \tilde{S}'} = \sum_i \sum_k \Lambda_{ik} \delta_{ik} \tilde{S}'_{ki} = \sum_i \Lambda_{ii} \tilde{S}'_{ii}.
\end{equation}

\FloatBarrier

\section{Shape of the defect orbital and the CBM in MgO with an oxygen vacancy}

\begin{figure}[H]
    \centering
    \includegraphics[width=0.75\linewidth]{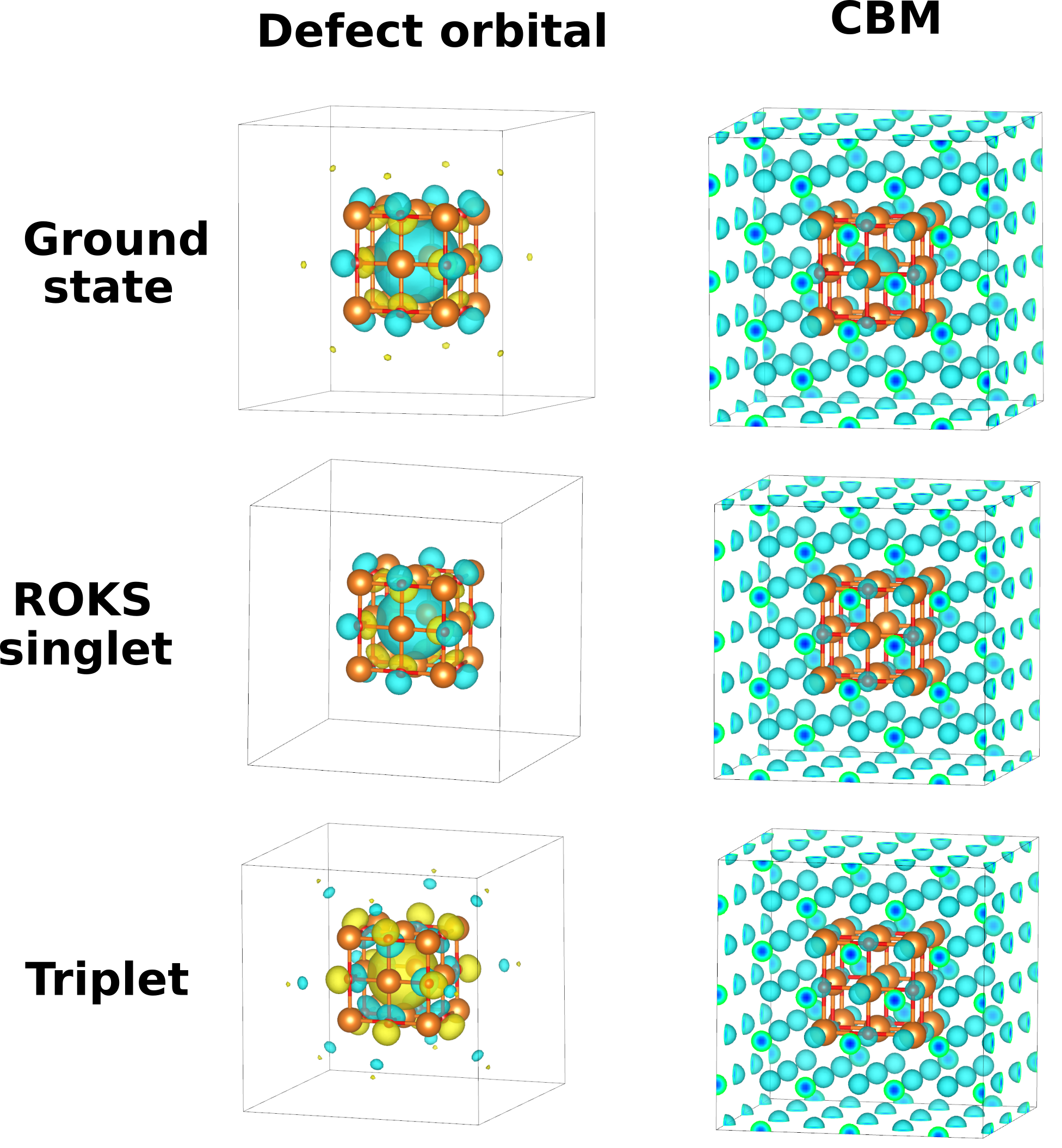}
    \caption{The shape of the localized defect orbital (left) and the conduction band minimum (right) of the $F^0$-center in MgO obtained from a ground state calculation (top), a ROKS calculation of the excited singlet state (middle), calculation of the triplet state (bottom). For better visibility only the atoms around the vacant site are shown.}
    \label{fig:rad_dist_3T1u}
\end{figure}

\clearpage

\section{Radial Distribution Functions around the oxygen vacancy in $\text{MgO}$}

\begin{figure}[H]
    \centering
    \includegraphics[width=0.75\linewidth]{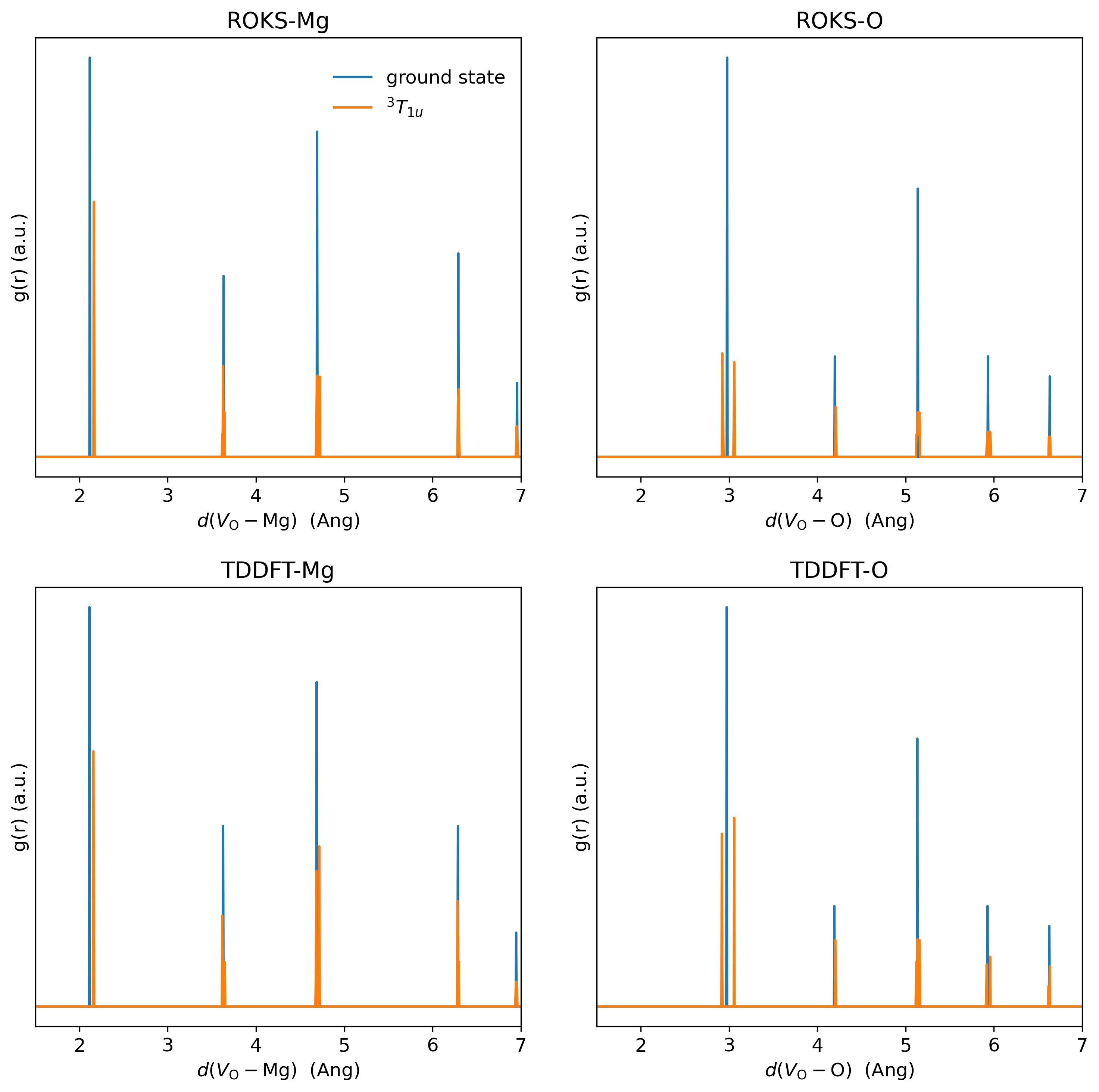}
    \caption{The radial distribution function around the oxygen vacancy in the excited state $^3T_{1u}$ (orange) and the groundstate (blue) at the respective equilibrium structures. The upper panels show results from ROKS-DDH for Mg (left) and O (right). The lower panels show the results obtained from TDDFT-DDH.}
    \label{fig:rad_dist_3T1u_ddh}
\end{figure}

\begin{figure}
    \centering
    \includegraphics[width=0.75\linewidth]{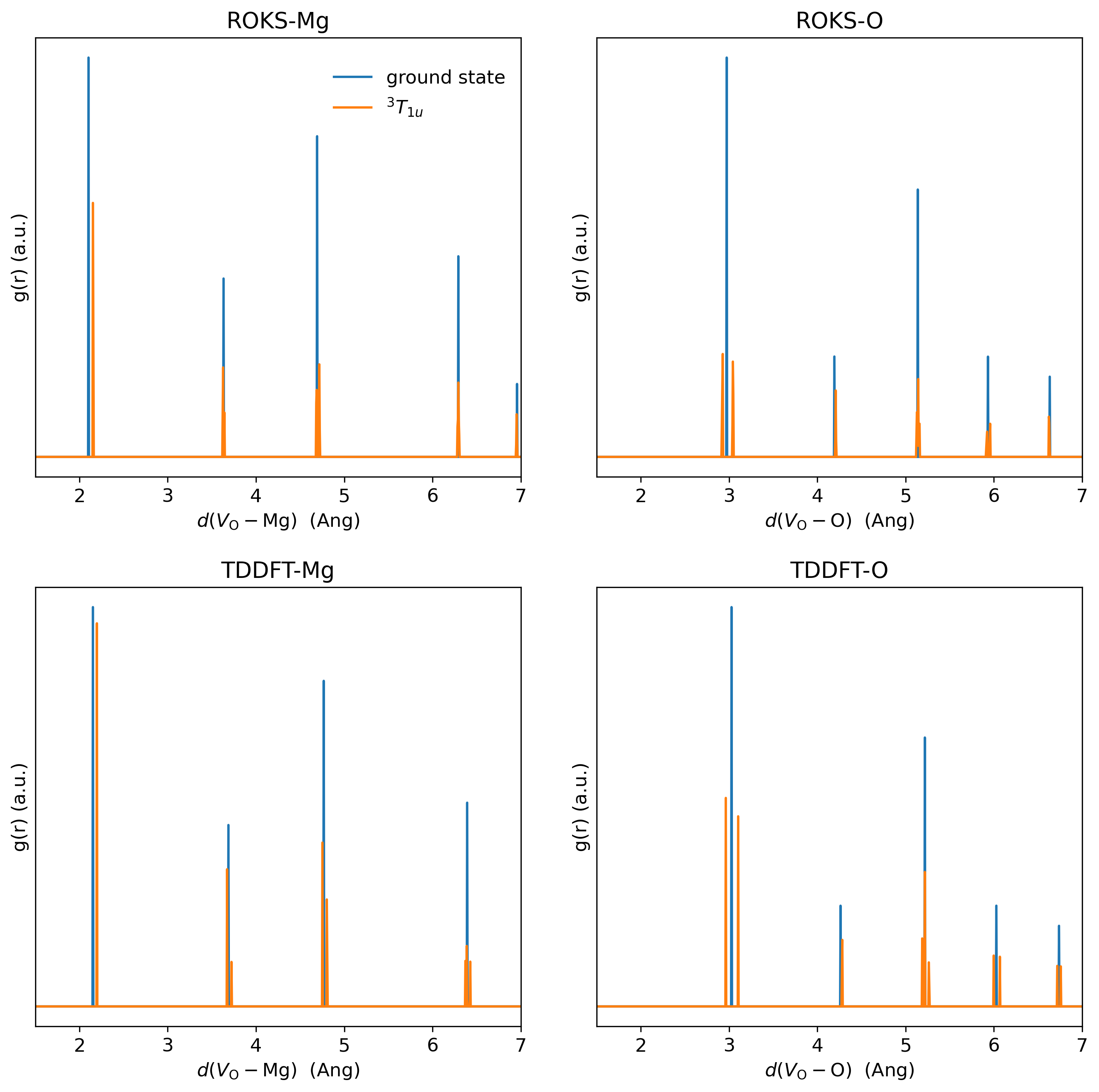}
    \caption{The radial distribution function around the oxygen vacancy in the excited state $^3T_{1u}$ (orange) and the groundstate (blue) at the respective equilibrium structures. The upper panels show results from ROKS-PBE for Mg (left) and O (right). The lower panels show the results obtained from TDDFT-PBE.}
    \label{fig:rad_dist_3T1u_pbe}
\end{figure}

\begin{figure}
    \centering
    \includegraphics[width=0.75\linewidth]{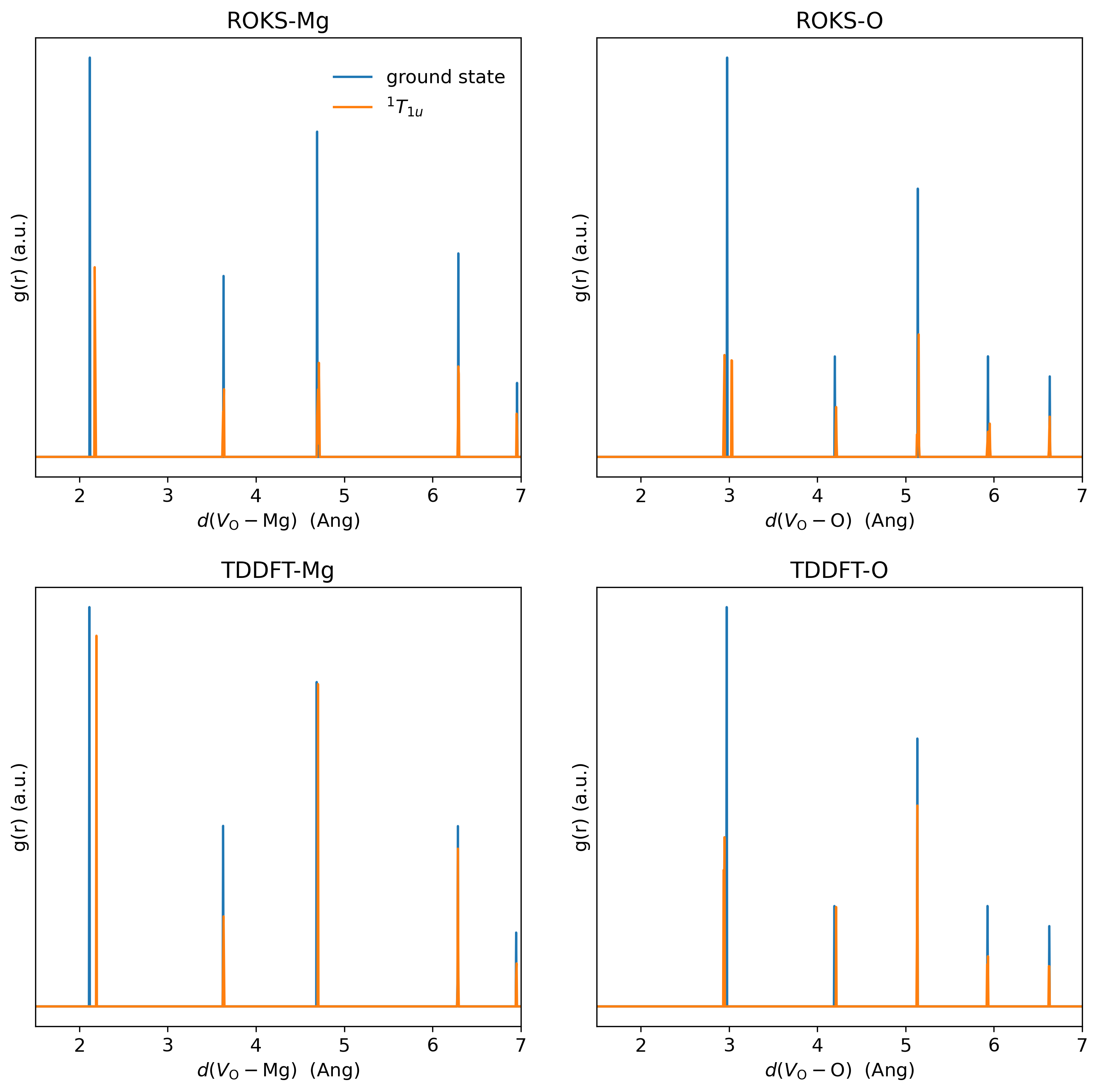}
    \caption{The radial distribution function around the oxygen vacancy in the excited state $^1T_{1u}$ (orange) and the groundstate (blue) at the respective equilibrium structures. The upper panels show results from ROKS-DDH for Mg (left) and O (right). The lower panels show the results obtained from TDDFT-DDH.}
    \label{fig:rad_dist_1T1u_ddh}
\end{figure}

\begin{figure}
    \centering
    \includegraphics[width=0.75\linewidth]{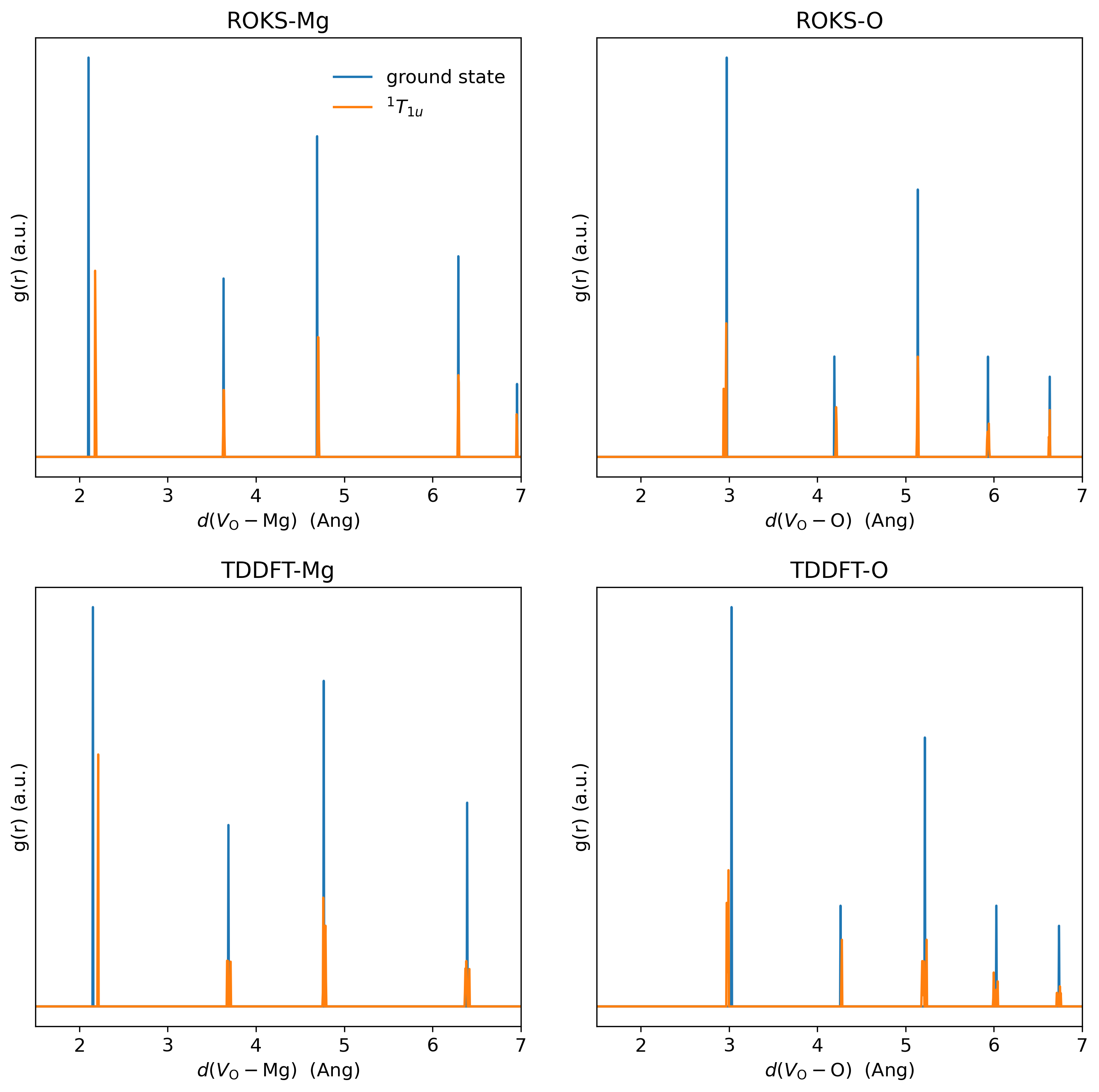}
    \caption{The radial distribution function around the oxygen vacancy in the excited state $^1T_{1u}$ (orange) and the groundstate (blue) at the respective equilibrium structures. The upper panels show results from ROKS-PBE for Mg (left) and O (right). The lower panels show the results obtained from TDDFT-PBE.}
    \label{fig:rad_dist_1T1u_pbe}
\end{figure}

\clearpage
\bibliographystyle{apsrev4-2}
\bibliography{ref}